\documentclass[onecolumn,superscriptaddress,amsmath,amssymb,floatfix,aps,notitlepage,showkeys,prx,longbibliography]{revtex4-1}
\usepackage{xr-hyper}
\usepackage{hyperref}
\usepackage{mathrsfs}

\usepackage{amssymb, amsbsy, amsmath, latexsym, dsfont, array, layout, graphics,mathrsfs,braket,amsfonts,amsthm, graphicx,subfigure, youngtab,color,bm,mathtools,braket,verbatim,url,cleveref,natbib,hypernat,extarrows,inputenc,multirow,bbm,enumitem}

\usepackage[usernames,dvipsnames]{xcolor}

\makeatletter
\newcommand*{\addFileDependency}[1]{
  \typeout{(#1)}
  \@addtofilelist{#1}
  \IfFileExists{#1}{}{\typeout{No file #1.}}
}
\makeatother

\newcommand*{\myexternaldocument}[1]{%
    \externaldocument{#1}%
    \addFileDependency{#1.tex}%
    \addFileDependency{#1.aux}%
}

\myexternaldocument{supple}

\newcommand{\mbf}[1]{\mathbf{#1}}

\definecolor{AtomicTangerine}{rgb}{1.0, 0.6, 0.4}

\makeatletter
\newcommand*{\inlineequation}[2][]{%
  \begingroup
    \refstepcounter{equation}%
    \ifx\\#1\\%
    \else
      \label{#1}%
    \fi
    \relpenalty=10000 %
    \binoppenalty=10000 %
    \ensuremath{%
      #2%
    }%
    ~\@eqnnum
  \endgroup
}
\makeatother

\newcolumntype{M}[1]{>{\centering\arraybackslash}m{#1}}
\newcolumntype{N}{@{}m{0pt}@{}}

\allowdisplaybreaks
\newcommand{\splitatcommas}[1]{%
  \begingroup
  \begingroup\lccode`~=`, \lowercase{\endgroup
    \edef~{\mathchar\the\mathcode`, \penalty0 \noexpand\hspace{0pt plus 1em}}%
  }\mathcode`,="8000 #1%
  \endgroup
}

\begin{document}
\title{Statistical mechanics of nanotubes}
\author{Siddhartha Sarkar}
\email{sarkarsi@umich.edu}
\affiliation{Department of Physics, University of Michigan, Ann Arbor, MI 48109, USA}
\affiliation{Department of Mechanical and Aerospace Engineering, Princeton University, Princeton, NJ 08544, USA}
\author{Mohamed El Hedi Bahri}
\email{mbahri@princeton.edu}
\affiliation{Department of Mechanical and Aerospace Engineering, Princeton University, Princeton, NJ 08544, USA}
\author{Andrej Ko\v{s}mrlj}
\email{andrej@princeton.edu}
\affiliation{Department of Mechanical and Aerospace Engineering, Princeton University, Princeton, NJ 08544, USA}
\affiliation{Princeton Institute for the Science and Technology of Materials, Princeton University, Princeton, NJ 08544, USA}

\begin{abstract}
We investigate the effect of thermal fluctuations on the mechanical properties of nanotubes by employing tools from statistical physics. For 2D sheets it was previously shown that thermal fluctuations effectively renormalize elastic moduli beyond a characteristic temperature-dependent thermal length scale (a few nanometers for graphene at room temperature), where the bending rigidity increases, while the in-plane elastic moduli reduce in a scale-dependent fashion with universal power law exponents. However, the curvature of nanotubes produces new phenomena. In nanotubes, competition between stretching and bending costs associated with radial fluctuations introduces a characteristic elastic length scale, which is proportional to the geometric mean of the radius and effective thickness. Beyond elastic length scale, we find that the in-plane elastic moduli stop renormalizing in the axial direction, while they continue to renormalize in the circumferential direction beyond the elastic length scale albeit with different universal exponents. The bending rigidity, however, stops renormalizing in the circumferential direction at the elastic length scale. These results were verified using molecular dynamics simulations.
\end{abstract}
\maketitle
\section{Introduction}
Atomically thin membranes have been a subject of interest over the last few decades~\cite{novoselov2005two,Ijima,MoS2,hBN,WS2} for their promising electronic and mechanical properties. Thin solid membranes are also ubiquitous in soft condensed matter~\cite{softmatter1,softmatter2,softmatter3} and biological systems~\cite{biology1,biology2,biology3,biology4,biology5,biology6,biology7,biology8,biology9}. In these contexts, the statistical mechanics of freely suspended elastic membranes have been studied extensively~\cite{nelson1987fluctuations,kantor1987crumpling,Paczuski,David1988,guitter1989thermodynamical,Aronovitz1989,aronovitz1988fluctuations,radzihovsky1991statistical,radzihovskySCSA,radzihovskySCSA2018,Mouhanna2009,Hasselman2010,Hasselman2011,AndrejWarped}. Theoretical studies of these membranes suggest that due to long-range interaction between local Gaussian curvatures mediated by in-plane phonons, arbitrarily large elastic membranes can remain flat at low enough temperatures~\cite{nelson1987fluctuations,kantor1987crumpling,Paczuski,David1988,guitter1989thermodynamical,Aronovitz1989}. In such membranes, the thermal fluctuations stiffen the bending rigidity $\kappa$ in a scale ($\ell$) dependent fashion,  $\kappa(\ell)\sim \ell^\eta$, and reduce the Young's modulus $Y(\ell)\sim \ell^{-\eta_u}$ beyond a temperature $T$ dependent length $\ell_\text{th} \sim \sqrt{\kappa_0^2/(k_B T Y_0)}$, where $\kappa_0$ and $Y_0$ are the microscopic bending rigidity and Young's modulus respectively. This result has been obtained using various methods such as perturbative renormalization group with $\epsilon$-expansion~\cite{guitter1989thermodynamical,aronovitz1988fluctuations,radzihovsky1991statistical}, $1/d$-expansion~\cite{Aronovitz1989}, self consistent screening approximation~\cite{radzihovskySCSA,radzihovskySCSA2018} and nonperturbative renormalization group~\cite{Mouhanna2009,Hasselman2010,Hasselman2011}, verified with numerical simulations~\cite{numerical1,BowickNumerics,MORSHEDIFARD2021104296}, and being studied in experiments~\cite{experimental1}.

Atomically thin cylindrical shell-like structures are important in the context of carbon nanotubes. Since its invention in 1991~\cite{Ijima}, carbon nanotubes have gained significant research interest due to their electronic properties~\cite{Tombler2000769}, high tensile strength~\cite{CNTtensilestrentgh,CNTMech&Therm}, thermal conductivity~\cite{CNTMech&Therm} and their ability to adsorb gases~\cite{CNTGasAbsorption}. They have been used in field-effect transistors~\cite{CNTFET}, composite materials~\cite{CNTComposite}, environmental monitoring~\cite{CNTGasAbsorption} etc. For these applications, it is important to study the effect of thermal fluctuations on the mechanical properties of nanotubes. While thermal fluctuations of flat sheets are well understood, much less is known about the response of nanotubes to thermal fluctuations.

Here we investigate the statistical mechanics of thin single-walled nanotubes at low temperatures within shallow shell approximation~\cite{SandersShell,Donnell,mushtari1961non}. Due to the presence of the curvature in nanotubes, the radial fluctuations along the axial direction cost both bending and stretching energy, whereas the radial fluctuations along the circumferential direction only cost bending energy. This competition between stretching and bending costs associated with height fluctuations introduces a characteristic elastic length scale ($\ell_\text{el} \sim \sqrt{Rt}$)~\cite{AndrejSphericalShell,komura}, which is proportional to the geometric mean of the radius and effective thickness. In typical carbon naotubes, this length is $\ell_\text{el} \lesssim 3 \text{nm}$. We show that below this length, the bending rigidity and in-plane moduli scales the same way as for flat membranes mentioned above. As will be discussed in detail in this article, beyond the elastic length scale, the in-plane elastic moduli stop renormalizing in the axial direction, while they continue to renormalize in the circumferential direction beyond the elastic length scale albeit with different universal exponents. The bending rigidity, however, stops renormalizing in the circumferential direction at the elastic length scale. We verify our theoretical findings with molecular dynamics simulations.

The remainder of the paper is organized as follows. In Sec.~\ref{sec:ElasticEnergy}, we review shallow shell theory description for cylindrical shells. In Sec.~\ref{sec:ThermalFluctuations}, we set up the statistical mechanics problem. In Sec.~\ref{sec:Scaling}, we perform renormalization group and scaling analysis to show how the elastic moduli scale in different regimes. In Sec.~\ref{sec:MD}, we compare the theoretically predictions with molecular dynamics simulations. In Sec.~\ref{sec:Conclusion}, we give concluding remarks and comment on possible further investigations for better understanding of the mechanical properties of thermalized cylindrical shells.

\section{Elastic energy of deformation}\label{sec:ElasticEnergy}
\begin{figure}[h!]
  \centering
  \includegraphics{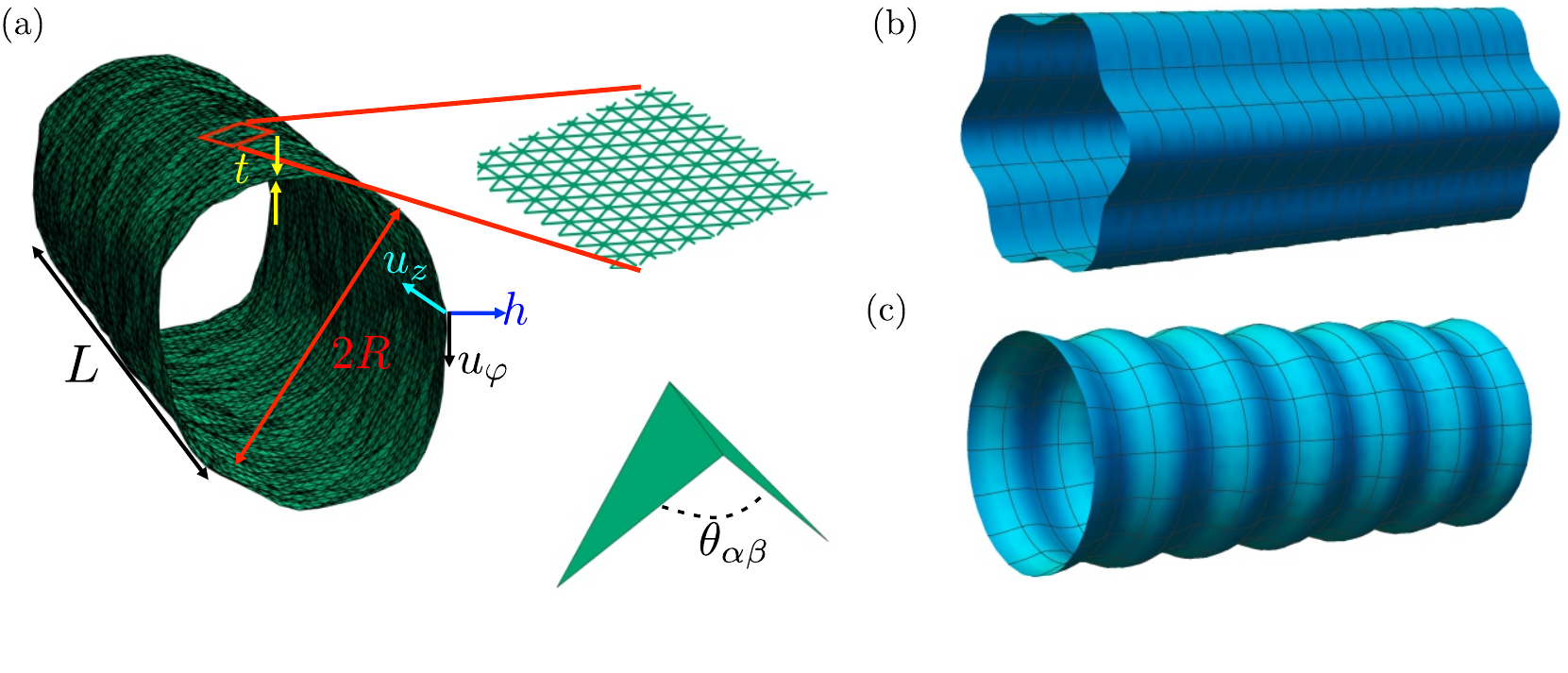}
  \caption{(a)~A snaphot of a thermally fluctuating cylindrical shell. The length $L$, diameter $2R$, thickness $t$, in-plane displacements $u_\varphi$ and $u_z$, radial displacement $h$ are shown. A small segment (shown in red box) on the shell is zoomed in to show that in simulations, the shell is modeled with triangular lattice of point masses connected by harmonic springs (green lines). At bottom right corner, the angle $\theta_{\alpha\beta}$ between two adjacent triangles of the triangular lattice is shown. (b)~A deformed cylindrical shell with only nonzero $\tilde{h}(q_\varphi \neq 0, q_z = 0)$ and $\tilde{h}(q_\varphi = 0, q_z = 0)$ does not cost stretching energy if the change in the length of circumference due to $\tilde{h}(q_\varphi \neq 0, q_z = 0)$ is compensated by $\tilde{h}(q_\varphi = 0, q_z = 0)$. (c)~A deformed cylindrical shell with nonzero $\tilde{h}(q_\varphi = 0, q_z \neq 0)$ and $\tilde{h}(q_\varphi = 0, q_z = 0)$ costs stretching energy different positions along the axis of the cylindrical shell have different circumference length.}
  \label{fig:figure1}
\end{figure}

The elastic energy of a deformed cylindrical shell can be estimated with shallow shell theory. To this end, let us consider a cylindrical shell with radius $R$ and length $L$ in its undeformed configuration. \footnote{In principle, a cylindrical shell should have different inner and outer radii since it can have finite thickness, for atomatically thin nanotubes thickness is not defined. However, as we will see later, an effective thickness can be defined in terms of its bending rigidity and in-plane moduli.} Then, any point on the undeformed shell can be written as $\bar{\mathbf{X}} = (R\cos\varphi,R\sin\varphi,z)$ in cartesian coordinates where the axis of the cylinder is in the $z$ direction. Since the radius $R$ is a constant, the shell can be parametrized by $(R\varphi,z)$ coordinates. The tangent vectors at any point on the shell can be written as $\bar{\mathbf{t}}_\varphi = \partial_\varphi \bar{\mathbf{X}} = (- \sin\varphi,\cos\varphi,0)$ and $\bar{\mathbf{t}}_z = \partial_z \bar{\mathbf{X}} = (0,0,1)$, whereas the normal is $\bar{\mathbf{N}} = (\cos\varphi,\sin\varphi,0)$. Note that here we used short-hand $\partial_\varphi \equiv \frac{1}{R}\frac{\partial}{\partial \varphi}$ and $\partial_z \equiv \frac{\partial}{\partial z}$. The reference metric is $\bar{g}_{ij} = \bar{\mathbf{t}}_i\cdot\bar{\mathbf{t}}_j = \delta_{ij}$ and curvature tensor is $\bar{b}_{ij} = \bar{\mathbf{N}}\cdot\partial_i \bar{\mathbf{t}}_j = -\frac{1}{R}\delta_{ij}\delta_{i\varphi}$ where $i \in \{\phi,z\}$. Then, in the deformed configuration, the displacement of each point can be decomposed into tangential components $u_i(R\varphi,z)$ (where $i \in \{\varphi,z\}$) and radial component $h(R\varphi,z)$ (see Fig.~\ref{fig:figure1}(a)) such that the coordinates of the points in deformed configuration are given by $\mathbf{X} = \bar{\mathbf{X}} + \mathbf{u}_\varphi \bar{\mathbf{t}}_\varphi + \mathbf{u}_z \bar{\mathbf{t}}_z + h \bar{\mathbf{N}}$. The tangent vectors in the deformed configuration are $\mathbf{t}_\varphi = \partial_\varphi \mathbf{X} = \bar{\mathbf{t}}_\varphi + \bar{\mathbf{t}}_\varphi\partial_\varphi u_\varphi-\frac{1}{R} u_\varphi\bar{\mathbf{N}}+\bar{\mathbf{t}}_z\partial_\varphi u_z+\bar{\mathbf{N}}\partial_\varphi h + \frac{1}{R}h\bar{\mathbf{t}}_\varphi$ and $\mathbf{t}_z = \partial_z \bar{\mathbf{X}} = \bar{\mathbf{t}}_z + \bar{\mathbf{t}}_\varphi \partial_z\mathbf{u}_\varphi + \bar{\mathbf{t}}_z \partial_z \mathbf{u}_z + \bar{\mathbf{N}} \partial_z h$. Then, the metric and the curvature tensors in the deformed configuration are:
\begin{equation}
\label{eq:MetricCurvature}
\begin{split}
    g_{ij} &\approx \delta_{ij} + \partial_i u_j +\partial_j u_i + (\partial_i h)(\partial_j h)+ \frac{2}{R}h \delta_{ij}\delta_{i\varphi},\\
    b_{ij} &\approx -\frac{1}{R}\delta_{ij}\delta_{i\varphi}+\partial_i\partial_j h,
\end{split}
\end{equation}
where we kept only the relevant terms (in the sense of Wilsonian renormalization~\cite{peskin2018introduction}) in displacements and their derivatives. This is termed as the Donnell-Mushtari-Vlasov approximation~\cite{Donnell,SandersShell,mushtari1961non}. The in-plane strain and bending strain tensors are defined as $u_{ij} \equiv (g_{ij}-\bar{g}_{ij})/2$ and $K_{ij} \equiv (b_{ij}-\bar{b}_{ij})$, respectively~\cite{SandersShell}. Then, from Eq.~\ref{eq:MetricCurvature}, we find
\begin{equation}
\label{eq:Strains}
\begin{split}
    u_{ij} &= (\partial_i u_j+\partial_j u_i+\partial_i h\partial_j h)/2 + \frac{h}{R}\delta_{ij}\delta_{i\varphi},\\
    K_{ij} &= \partial_i\partial_j h,
\end{split}
\end{equation}
where $i,j \in \{\varphi, z\}$. Notice that first term inside the parenthesis in the expression of in-plane strain $u_{ij}$ is the same as in flat sheets. However, the second term is not present for flat sheets and will couple in-plane strains with the radial undulation field $h$ due to curvature $1/R$. The free energy cost of the shell deformation is given by:
\begin{equation}
\label{eq:FreeEnergy}
    F = \int_A dA\, \frac{1}{2}\left[\kappa_0(K_{ii})^2 + \lambda_0 (u_{ii})^2 + 2\mu_0 (u_{ij})^2\right],
\end{equation}
where the first term accounts for the bending energy and the last two terms are the in-plane stretching energy. Note that we did not consider the term consisting of the Gaussian curvature because we are interested in studying the system under periodic boundary conditions, where the integral of the Gaussian curvature is zero due to the Gauss-Bonnet theorem. Here, $\kappa_0$ is the microscopic (bare) bending rigidity of the material and has dimensions of energy, whereas $\lambda_0$ and $\mu_0$ are the microscopic (bare) Lam\'e coefficients and has the dimension of energy per area and $dA = d\varphi dz R$ is the infinitesimal area element with $A = 2\pi R L$ being the area of the system. We furthermore define $ds$ as the infinitesimal line element at the boundary, $\partial A$ is the boundary of the cylinder, $\mathbf{T}$ is the traction force at the boundary. The components of the traction force can be written as $T_i = n_j \sigma_{ij}$ where $\sigma_{ij}$ is the homogeneous stress tensor and $n_j$ is the in-plane normal to at the boundary. Replacing this expression of the traction force $T_i$ and using Stokes' theorem we get:
\begin{equation}
\begin{split}
    &\int_{\partial A} ds\, T_i u_i = \int_{\partial A} ds\, n_j \sigma_{ij} u_i = \int_A dA \partial_j (\sigma_{ij} u_i) = \int_A dA \sigma_{ij} \partial_j u_i = \sigma_{ij} \int_A dA\frac{1}{2}(\partial_j u_i+\partial_i u_j),\\
    &F = \int_A dA\, \frac{1}{2}\left[\kappa_0(\nabla^2 h)^2 + \lambda_0 (u_{ii})^2 + 2\mu_0 (u_{ij})^2-\sigma_{ij} (\partial_j u_i+\partial_i u_j)\right],
\end{split}
\end{equation}
where we used the facts that $\sigma_{ij} = \sigma_{ji}$ and $K_{ii} = \nabla^2 h$. 
We also assume that the microscopic material is isotropic which is reasonable for carbon nanotubes. However, inspection of the last term in the expression of $u_{ij}$ in Eq.~\ref{eq:Strains} shows that it is anisotropic. This means that as we integrate small wavelength degrees of freedom to investigate long wavelength behavior, the moduli may become anisotropic. Keeping this in mind, we write the more general form of the free energy is given below:
\begin{equation}
    F = \int_A dA\, \frac{1}{2}\left[B_{ijkl}^0(\partial_i\partial_j h)(\partial_k \partial_l h) + C_{ijkl}^0 u_{ij}u_{kl}-\sigma_{ij} (\partial_j u_i+\partial_i u_j)\right],
\end{equation}
where $B_{ijkl}^0$ and $C_{ijkl}^0$ are the most general bare bending rigidity and in-plane elastic tensors, respectively. However, because of the mirror planes $x-y$ and $r-z$, the anisotropy can be at most orthorhombic, meaning that $B_{\varphi\varphi\varphi z} = B_{\varphi z z z} = C_{\varphi\varphi\varphi z} = C_{\varphi z z z} = 0$. This along with the major and minor symmetries of $B_{ijkl}$ and $C_{ijkl}$ tensors mean that the only independent moduli are $B_{\varphi\varphi\varphi\varphi}$, $B_{\varphi\varphi z z}$, $B_{zzzz}$, $C_{\varphi\varphi\varphi\varphi}$, $C_{\varphi\varphi z z}$, $C_{\varphi z \varphi z}$ and $C_{zzzz}$. For isotropic bare rigidities, $B_{ijkl}^0 = \kappa_0 \delta_{ij}\delta_{kl}$ and $C_{ijkl}^0 = \lambda_0 \delta_{ij}\delta_{kl} + \mu_0 (\delta_{ik}\delta_{jl}+\delta_{il}\delta_{jk})$.
\section{Thermal fluctuations}\label{sec:ThermalFluctuations}
The effect of thermal fluctuations can be seen in the correlation functions obtained from the functional integrals:
\begin{subequations}
\begin{equation}
\label{eq:HeightZeroMode}
    \langle h \rangle \equiv \langle h(\mathbf{x}) \rangle = \frac{1}{Z} \int \mathcal{D}[u_i,h] h(\mathbf{x}) e^{-F/k_B T},\\
\end{equation}
\begin{equation}
\label{eq:HeightCorrelation}
    G_{hh}(\mathbf{x}_2-\mathbf{x}_1) \equiv \langle \delta h(\mathbf{x}_2) \delta h(\mathbf{x}_1) \rangle = \frac{1}{Z} \int \mathcal{D}[u_i,h] \delta h(\mathbf{x}_2) \delta h(\mathbf{x}_1) e^{-F/k_B T},\\
\end{equation}
\begin{equation}
\label{eq:InPlaneCorrelation}
    G_{u_iu_j}(\mathbf{x}_2-\mathbf{x}_1) \equiv \langle u_i(\mathbf{x}_2) u_j(\mathbf{x}_1) \rangle = \frac{1}{Z} \int \mathcal{D}[u_i,h] u_i(\mathbf{x}_2) u_j(\mathbf{x}_1) e^{-F/k_B T},\\
\end{equation}
\begin{equation}
    Z = \int \mathcal{D}[u_i,h] e^{-F/k_B T},\\
\end{equation}
\end{subequations}
where $T$ is the temperature, $k_B$ is the Boltzmann's constant, $Z$ is the partition function, $\delta h(\mathbf{x}) = h(\mathbf{x}) - \langle h \rangle$, $\mathbf{x} \equiv (R\varphi,z)$, and in Eqs.~\ref{eq:HeightZeroMode},~\ref{eq:HeightCorrelation} and~\ref{eq:InPlaneCorrelation} , we used the fact the system is translationally invariant.

In the following, we decompose the radial displacement field as $h(\mathbf{x}) = h_0 + \tilde{h}(\mathbf{x})$, where $h_0 = (1/A) \int dA\, h(\mathbf{x})$ is the homogeneous part of the undulation field, and $(1/A) \int dA\, \tilde{h}(\mathbf{x}) = 0$. Then, $\langle h_0 \rangle = (1/A) \int dA \langle h(\mathbf{x}) \rangle = \langle h \rangle$. With this knowledge then, $\delta h(\mathbf{x}) = \tilde{h}(\mathbf{x}) + h_0 - \langle h_0 \rangle$. Similarly, the in-plane strain fields can be decomposed into $u_{ij}(\mathbf{x}) = u_{ij}^0+\tilde{u}_{ij}(\mathbf{x})$. However, the homogeneous strain $u_{\varphi\varphi}^0$ and the zero mode of the radial fluctuation $h_0$ are related by $u_{\varphi\varphi}^0 = h_0/R$ because increasing the length in azimuthal direction effectively increases the radius. Integrating out the homogeneous fields $h_0$ and $u_{ij}^0$, we obtain the following effective free energy:
\begin{equation}
\label{eq:EffectiveEnergy1}
\begin{split}
    F_{\text{eff},1} &= -k_B T \ln\left(\int \mathcal{D}[u_{\varphi z}^0,h_0,u_{zz}^0]e^{-F/k_BT}\right)\\
    &= \int_A dA\, \frac{1}{2}\left[B_{ijkl}^0(\partial_i\partial_j \tilde{h})(\partial_k \partial_l \tilde{h}) + C_{ijkl}^0 \tilde{u}_{ij}\tilde{u}_{kl}+\sigma_{ij} (\partial_i \tilde{h})(\partial_j \tilde{h})\right]\\
    &= F_0 + F_I,\\
    F_0/A &= \int_A dA\, \frac{1}{2}\left[B_{ijkl}^0(\partial_i\partial_j \tilde{h})(\partial_k \partial_l \tilde{h}) + \frac{C_{\varphi\varphi\varphi\varphi}^0}{R^2}\tilde{h}^2+\sigma_{ij} (\partial_i \tilde{h})(\partial_j \tilde{h})\right]\\ 
    &\phantom{=}+\int_A dA\, \frac{1}{2}\left[C_{ijkl}^0(\partial_i\tilde{u}_j)(\partial_k \tilde{u}_l)+C_{ij\varphi\varphi}^0(\partial_i\tilde{u}_j)\frac{\tilde{h}}{R}\right],\\
    F_I/A &= \int_A dA\,\frac{1}{2}\left[C_{ijkl}^0\partial_i \tilde{u}_j (\partial_k \tilde{h})(\partial_l \tilde{h}) + C_{\varphi\varphi kl}^0\frac{\tilde{h}}{R}(\partial_k \tilde{h})(\partial_l \tilde{h})\right]\\
    &\phantom{=} +\int_A dA\, \frac{1}{8} C_{ijkl}^0 (\partial_i \tilde{h})(\partial_j \tilde{h}) (\partial_k \tilde{h})(\partial_l \tilde{h}),
\end{split}
\end{equation}
where $F_0$ and $F_I$ are the harmonic and anharmonic parts of the effective free energy $F_{\text{eff},1}$. A similar functional integral shows that the average radial and axial extensions are:
\begin{subequations}
\begin{equation}
\label{eq:RadialShrink}
    \frac{\langle h \rangle}{R} =  \frac{\langle h_0 \rangle}{R} =  \frac{\sigma_{\varphi\varphi}}{Y_{\varphi\varphi}^0} - \frac{\nu_{\varphi\varphi}^0\sigma_{zz}}{Y_{\varphi\varphi}^0} - \frac{1}{2} \langle (\partial_\varphi \tilde{h})^2\rangle,
\end{equation}
\begin{equation}
\label{eq:AxialShrink}
    \frac{\langle \Delta L \rangle}{L} = \langle u_{zz}^0 \rangle =  \frac{\sigma_{zz}}{Y_{zz}^0} - \frac{\nu_{zz}^0\sigma_{\varphi\varphi}}{Y_{zz}^0} - \frac{1}{2} \langle (\partial_z \tilde{h})^2\rangle,
\end{equation}
\end{subequations}
where $Y_{\varphi\varphi}^0 = C_{\varphi\varphi\varphi\varphi}- C_{\varphi\varphi zz}^2/C_{zzzz}$ and $Y_{zz}^0 = C_{zzzz}- C_{\varphi\varphi zz}^2/C_{\varphi\varphi\varphi\varphi}$ are the 2-dimensional Young's moduli in the azimuthal and axial directions respectively, and $\nu_{\varphi\varphi} = C_{\varphi\varphi zz}/C_{zzzz}$ and $\nu_{zz} = C_{\varphi\varphi zz}/C_{\varphi\varphi\varphi\varphi}$ are the Poisson's ratios in azimuthal and axial directions respectively. For isotropic microscopic properties, $Y_{\varphi\varphi}^0 = Y_{zz}^0 = 4\mu_0 (\lambda_0 + \mu_0)/(\lambda_0+2\mu_0) \equiv Y_0$,
and $\nu_{\varphi\varphi}^0 = \nu_{zz}^0 = \lambda_0/(\lambda_0+2\mu_0)\equiv \nu^0_\text{iso}$. The last terms in Eqs.~\ref{eq:RadialShrink} and~\ref{eq:AxialShrink} are negative meaning that in absence of any normal stress ($\sigma_{\varphi\varphi} = \sigma_{zz} = 0$), the radius and the length of the cylindrical shell shrink under thermal fluctuations. This is because nonuniform radial fluctuations $\tilde{h}(\mathbf{x})$ at fixed radius would increase the integrated area, with a large stretching energy cost. The system prefers to wrinkle and shrink its radius to gain entropy while keeping the integrated area of the convoluted shell approximately constant.

For renormalization group calculations, it is sometimes helpful to integrate out the in-plane phonons and write an effective free energy as a functional of radial undulations:
\begin{equation}
\label{eq:EffectiveEnergy}
\begin{split}
    F_\text{eff} &= -k_B T \ln \left(\int \mathcal{D}[u_\varphi,u_z]e^{-F_{\text{eff},1}/k_BT}\right) = F_\text{eff}^0 + F_\text{eff}^I,\\
    F_\text{eff}^0/A &= \sum_{\mathbf{q}\neq \mathbf{0}} \frac{1}{2} \left[B_{ijkl}^0 q_iq_jq_kq_l + \frac{N(C_{ijkl}^0) q_z^4}{R^2D(C_{ijkl}^0;\mathbf{q})}+\sigma_{ij}q_iq_j\right] \tilde{h}(\mathbf{q})\tilde{h}(-\mathbf{q}),\\
    F_\text{eff}^I/A &= \sum_{\mathbf{q}_1+\mathbf{q}_2 = - \mathbf{q} \neq \mathbf{0}} \frac{q_z^2}{2q^2}[q_{1i}P^T_{ij}(\mathbf{q})q_{2j}]\frac{N(C_{ijkl}^0) q^4}{RD(C_{ijkl}^0;\mathbf{q})}\tilde{h}(\mathbf{q})\tilde{h}(\mathbf{q}_1)\tilde{h}(\mathbf{q}_2)\\
    &\phantom{=}+ \sum_{\substack{\mathbf{q}_1+\mathbf{q}_2 = \mathbf{q} \neq \mathbf{0}\\\mathbf{q}_3+\mathbf{q}_4 = -\mathbf{q} \neq \mathbf{0}}} \frac{1}{8}[q_{1i}P^T_{ij}(\mathbf{q})q_{2j}][q_{3i}P^T_{ij}(\mathbf{q})q_{4j}]\frac{N(C_{ijkl}^0) q^4}{D(C_{ijkl}^0;\mathbf{q})}\tilde{h}(\mathbf{q}_1)\tilde{h}(\mathbf{q}_2)\tilde{h}(\mathbf{q}_3)\tilde{h}(\mathbf{q}_4),\\
    N(C_{ijkl}) &= \det(C_{ijkl})/4 =  C_{\varphi\varphi\varphi\varphi}C_{\varphi z \varphi z}C_{zzzz} - C_{\varphi\varphi z z}^2C_{\varphi z \varphi z},\\
    D(C_{ijkl};\mathbf{q}) &= \det(C_{kilj}q_k q_l)=\frac{1}{2}(\varepsilon_{i_1 i_2}\varepsilon_{j_1 j_2} C_{k_1 i_1 l_1 j_1}C_{k_2 i_2 l_2 j_2}q_{k_1}q_{k_2}q_{l_1}q_{l_2})\\
    &= C_{\varphi\varphi\varphi\varphi}C_{\varphi z \varphi z} q_\varphi^4 + (C_{\varphi\varphi\varphi\varphi}C_{zzzz} - 2C_{\varphi\varphi z z}C_{\varphi z\varphi z}-C_{\varphi\varphi z z}^2)q_\varphi^2 q_z^2\\
    &\phantom{=} + C_{zzzz}C_{\varphi z \varphi z} q_z^4,
\end{split}
\end{equation}
where $P_{ij}^T(\mathbf{q}) = (\delta_{ij}-q_i q_j/q^2)$, $\varepsilon_{ij}$ is the permutation symbol, and we took Fourier transform of the radial undulation field $\tilde{h}(\mathbf{q}) = \int_A (dA/A) \tilde{h}(\mathbf{x}) e^{-i\mathbf{q}\cdot\mathbf{x}}$. Note that in the isotropic case, $N q^4/ D(C_{ijkl}^0;\mathbf{q}) = Y_0$. Note that $F_\text{eff}^0$ and $F_\text{eff}^I$ are the harmonic and anharmonic part of the effective free energy $F_\text{eff}$. Then, within harmonic approximation one can read off the Fourier transform $G_{hh}^0(\mathbf{q}) = \int_A (dA/A) G_{hh}^0(\mathbf{x}) e^{-i\mathbf{q}\cdot\mathbf{x}}$ of the correlation function $G_{hh}^0(\mathbf{x})$ (the superscript ``$0$'' is for harmonic approximation):
\begin{equation}
\label{eq:GreenFunctionHarmonic}
    G_{hh}^0(\mathbf{q}) \equiv \langle |\tilde{h}(\mathbf{q})|^2 \rangle_0 = \frac{k_B T/A}{B_{ijkl}^0 q_i q_j q_k q_l + \frac{N(C_{ijkl}^0) q_z^4}{R^2D(C_{ijkl}^0;\mathbf{q})}+\sigma_{ij}q_iq_j} \stackrel{\text{isotropic}}{=} \frac{k_B T/A}{\kappa_0 q^4 + \frac{Y_0 q_z^4}{R^2 q^4}+\sigma_{ij}q_iq_j},
\end{equation}
where  $\langle \rangle_0$ is the harmonic average. The effect of the anharmonic terms is to replace the bare parameters $B_{ijkl}^0$, $C_{ijkl}^0$ and $\sigma_{ij}$ with scale dependent renormalized parameters $B_{ijkl}^R(\mathbf{q})$, $C_{ijkl}^R(\mathbf{q})$ and $\sigma_{ij}^R(\mathbf{q})$:
\begin{equation}
\label{eq:GreenFunctionRN}
\begin{split}
    G_{hh}(\mathbf{q}) &\equiv \langle |\tilde{h}(\mathbf{q})|^2 \rangle = \frac{k_B T/A}{B_{ijkl}^R(\mathbf{q}) q_i q_j q_k q_l + \frac{N(C^R_{ijkl}(\mbf{q})) q_z^4}{R^2D(C^R_{ijkl}(\mbf{q});\mbf{q})}+\sigma_{ij}^R(\mathbf{q}) q_iq_j}.\\
\end{split}
\end{equation}
Before going into the details of renormalization, it is useful to gain some insights from the Green's function in Eq.~\ref{eq:GreenFunctionHarmonic}. In the limit $R \rightarrow \infty$, it gives back the Green's function for isotropic sheets $G_{hh}^{s,0}(\mathbf{q}) = \frac{k_B T/A}{\kappa_0 q^4 +\sigma_{ij}q_iq_j}$. Because of the presence of anisotropic curvature in the cylindrical shell, we have, in the denominator, an extra direction dependent term $(Y_0 q_z^4)/(R^2 q^4)$ which suppresses the amplitude of the radial fluctuations in the axial direction. This because the Fourier modes of radial fluctuations which are in axial direction necessarily cost stretching energy along with bending energy, whereas the Fourier modes of radial fluctuations which are in azimuthal direction only cost bending energy (see Fig.~\ref{fig:figure1}(b) and (c))~\cite{komura}. Furthermore, setting external stresses to zero $\sigma_{ij} = 0$ and equating the two remaining terms in the denominator of Eq.~\ref{eq:GreenFunctionHarmonic}, we obtain a characteristic wave vector:
\begin{equation}
\label{eq:ElasticLengthHarmonic}
    q_\text{el}^0 \equiv \frac{\pi}{\ell_\text{el}^0} = \left(\frac{Y_0}{\kappa_0 R^2}\right)^\frac{1}{4} = \frac{\gamma^{\frac{1}{4}}}{R},
\end{equation}
where $\gamma = Y_0 R^2/\kappa_0$ is the F\"oppl-von Karman number. In the theory of shallow shells, $\kappa_0 \sim Et^3$ and $Y_0 \sim E t$, where $E$ is 3-dimensional Young's modulus of the material and $t$ is the thickness of shell. In atomically thin systems thickness is not well defined; however, we can define an effective thickness as $t \sim \sqrt{\kappa_0/Y_0}$ Then, the F\"oppl-von Karman number $\gamma \sim R^2/t^2$ meaning that the larger $\gamma$ is, the larger the radius $R$ is with respect to the thickness $t$. The length scale $\ell_\text{el}^0$ obtained from this is what we will call the ``elastic length scale.'' The superscript ``0'' is for harmonic approximation; we will see later that when we take renormalization of the parameters due to the anharmonic terms into account, the expression of $\ell_\text{el}$ changes slightly. As we will see in the next section, the effect of the curvature on the renormalization of the material parameters is negligible for scales smaller the elastic length scale $\ell_\text{el}$ and will be important at scales larger than this. Another important length scale that is important for both isotropic sheets and cylindrical shells comes from the form of the Green's function: $G_{hh}^0 \sim \frac{k_BT}{A\kappa_0 q^4}$ when $\sigma_{ij} = 0$ and $q\gg q_\text{el}^0$. Therefore, the largest amplitude of the radial (height in case of sheets) fluctuations occur when $q \sim 1/L$ and $A \sim L^2$ (where is the system size) giving largest amplitude of the radial fluctuation as $h_\text{th} \sim L \sqrt{k_B T/\kappa_0} = L \tau^{1/2}$ ($\tau \equiv k_B T/\kappa_0$ is the nondimensional temperature). Anharmonic terms become important when this amplitude is of the order of the thickness $h_\text{th} \sim t \sim \sqrt{\kappa_0/Y_0}$. This gives us a length $\ell_\text{th} \sim \sqrt{\kappa_0^2/(k_BTY_0)}$ called thermal length scale~\cite{AndrejRibbon} (we will get a better estimate of $\ell_\text{th}$ later in Eq.~\ref{eq:ElasticLength}). The effect of the anharmonic terms are only important when the system size is larger than the thermal length scale $\ell_\text{th}$. These two length scales $\ell_\text{th}$ and $\ell_\text{el}$ divides the scale dependence of the material parameters into three regimes. We will be interested in the limit where $\ell_\text{th} \ll \ell_\text{el}$ because as was discussed above, below $\ell_\text{th}$ the anharmonic terms are not important and thus the theory is trivial. Therefore, keeping this length smaller than other important length scales enables us to see all possible non-trivial scalings due to the anharmonic terms. In addition, we will be interested in zero external stress limit $\sigma_{ij}^R(q) = 0$.
\section{Renormalization group and scaling analysis}\label{sec:Scaling}
The effect of the anharmonic terms in Eq.~\ref{eq:EffectiveEnergy1} at a given scale $\ell^* \equiv \pi/q^*$ can be obtained by systematically integrating out all degrees of freedom on smaller scales (i.e., larger wave vectors). This can be done by splitting the displacement fields into pieces: $g_<(\mathbf{r}) = \sum_{|\mathbf{q}|<q^*} e^{i \mathbf{q} \cdot \mathbf{r}} g(\mathbf{q})$ and $g_>(\mathbf{r}) = \sum_{\Lambda>|\mathbf{q}|>q^*} e^{i \mathbf{q} \cdot \mathbf{r}} g(\mathbf{q})$, where $g \in \{\tilde{u}_i,\tilde{h}\}$ and $a \equiv \pi/\Lambda$ is a microscopic cutoff, and integrating out $g_>$ as
\begin{equation}
\label{eq:IntegrationFastModes}
\begin{split}
    F_{\text{eff},1}(\ell^*) &= -k_B T \ln\left(\int \mathcal{D}[\tilde{u}_{i,>},\tilde{h}_>]e^{-F_{\text{eff},1}/k_BT}\right),\\
    F_{\text{eff}}(\ell^*) &= -k_B T \ln\left(\int \mathcal{D}[\tilde{h}_>]e^{-F_{\text{eff},1}/k_BT}\right).
\end{split}
\end{equation}
\subsection{Scaling analysis for $\ell^* \ll \ell_\text{el}$}
\label{sec:IsotropicScaling}
\begin{itemize}
\item $\ell^* \ll \ell_\text{th} \ll \ell_\text{el}^0$: we have $C_{\varphi\varphi\varphi\varphi}^0 |\tilde{h}(\mbf{q})|^2/R^2 \ll  \kappa_0 (q^*)^4 |\tilde{h}(\mbf{q})|^2$, meaning that we can ignore the term $C_{\varphi\varphi\varphi\varphi}^0 \tilde{h}^2/R^2$ from Eq.~\ref{eq:EffectiveEnergy1}, and similarly we can ignore the term $N(C_{ijkl}^0) q_z^4/(R^2 D(C_{ijkl}^0;\mathbf{q}))\tilde{h}(\mbf{q})\tilde{h}(-\mathbf{q}) \sim Y_0 q_z^4/(R^2 q^4)\tilde{h}(\mbf{q})\tilde{h}(-\mathbf{q})$ from Eq.~\ref{eq:EffectiveEnergy}. Since $\ell^* \ll \ell_\text{th}$, the anharmonic terms are not important as was discussed in the last section. Therefore, the effective harmonic free energy is
\begin{equation}
    F_{\text{eff}}(\ell^*) = \sum_{\substack{\mathbf{q} \neq \mbf{0}\\ q < q^*}} \frac{A}{2} \left[B_{ijkl}^0 q_iq_jq_kq_l + \sigma_{ij}q_iq_j\right] = \sum_{\substack{\mathbf{q} \neq \mbf{0}\\ q < q^*}} \frac{A}{2} \left[\kappa_0 q^4 + \sigma_{ij}q_iq_j\right] \tilde{h}(\mathbf{q})\tilde{h}(-\mathbf{q}).
\end{equation}
This free energy is isotropic and the material parameters remain the same as their bare values. The na\"ive (Gaussian or harmonic) dimensions of the quantities $\tilde{h}(\mbf{q})$ and $\sigma_{ij}$ requiring that $[F_{\text{eff}}] = 0$ and $[\kappa_0] = 0$ and expressing dimensionality in wave-vector units~\cite{Amit}:
\begin{equation}
    [\tilde{h}(\mbf{q})] = -1 = (D-4)/2 \equiv -\zeta_h, [\sigma_{ij}] = 2,
\end{equation}
where $D=2$ is the dimension of the system. The meaning of these dimensions is that under the scale transformation $\mathbf{q} \rightarrow \mathbf{q}' = b\mathbf{q}$ or $\mathbf{x} \rightarrow \mathbf{x}' = b^{-1}\mathbf{x}$, if we transform $\tilde{h}(\mbf{q}) \rightarrow \tilde{h}'(\mbf{q}') = b^{[\tilde{h}(\mbf{q})]}\tilde{h}(b\mbf{q}) = b^{-1}\tilde{h}(b\mbf{q})$ and $\sigma_{ij} \rightarrow \sigma_{ij}' = b^{[\sigma_{ij}]}\sigma_{ij} = b^{2}\sigma_{ij}$, the harmonic theory remains invariant. With these dimensions we go back to $F_{\text{eff},1}$ in Eq.~\ref{eq:EffectiveEnergy1} and find the na\"ive dimensions of the following quantities:
\begin{equation}
    [C_{ijkl}^0] = 2 = 4-D, [\tilde{u}_i] = -1 = D-3 = 1-2\zeta_h, [1/R] = 1 = D/2.
\end{equation}
This means that under scale transformation $\mathbf{q} \rightarrow \mathbf{q}' = b\mathbf{q}$, $C_{ijkl}'^0 = b^{4-D}C_{ijkl}^0$ meaning as we zoom out of the system $C_{ijkl}'^0$ grows for dimensions $D<4$. Since $C_{ijkl}^0$ are the coefficients of the anharmonic terms in Eq.~\ref{eq:EffectiveEnergy1}, the anharmonic terms are important for dimensions $D \leq 4$. This implies that the upper critical dimension of the theory in Eq.~\ref{eq:EffectiveEnergy1} is $D_u = 4$ and the anharmonic terms are relevant in the physical dimension $D = 2$. This means that the anharmonic terms will renormalize the parameters $B_{ijkl}^0$ and $C_{ijkl}^0$ at least in the regime $\ell_\text{th} < \ell^* \ll \ell_\text{el}^0$ which we discuss next.
\item $\ell_\text{th} < \ell^* \ll \ell_\text{el}^0$: In this regime, the anharmonic renormalize the parameters $B_{ijkl}^0$ and $C_{ijkl}^0$ to parameters $B_{ijkl}^R(\mbf{q}^*)$ and $C_{ijkl}^R(\mbf{q}^*)$ at scale $q^*$. However, for $\ell^* \gtrsim \ell_\text{th}$ when the parameters are only mildly renormalized, the anisotropic term $N^R(\mbf{q}) q_z^4/(R^2 D^R(\mbf{q}))$ in the denominator of the Green's function in Eq.~\ref{eq:GreenFunctionRN} is still small w.r.t. the first term $B_{ijkl}^R(\mathbf{q})q_iq_jq_kq_l$(we will show the consistency of this once we obtain the renormalization group flow equations). Therefore, we can implement an isotropic (circular) momentum-shell renormalization group scheme. We first integrate out all Fourier modes in a thin momentum shell $\Lambda/b < q< \Lambda$ and $b \equiv \ell^*/a = e^s$, with $s \ll 1$. Next we rescale lengths and fields:
\begin{equation}
    \mathbf{r} = b \mathbf{r}',\, \tilde{h}(\mbf{r}) = b^{\zeta_h} \tilde{h}(\mbf{r}'),\, \tilde{u}_i(\mbf{r}) = b^{2\zeta_h-1} \tilde{u}_i(\mbf{r}')
\end{equation}
where $\zeta_h$ is the field-rescaling exponent. We find it convenient to work directly with a $D = 2$-dimensional cylindrical shell embedded in $d = 3$ space, rather than introducing an expansion in $\epsilon = 4 - D$~\cite{aronovitz1988fluctuations,guitter1989thermodynamical}. In principle, this is dangerous and can give wrong results because if we work in $D=2$, $\epsilon = 4-2 = 2$ is large and there is no small parameter in the perturbative renormalization scheme. However, later we show using the molecular dynamics simulations that the results obtained by directly working in $D=2$ matches with simulations. Finally, we define new elastic moduli $B_{ijkl}'$, $C_{ijkl}'$, and a new external pressure $\sigma_{ij}'$, such that the free-energy functional in Eqs.~\ref{eq:EffectiveEnergy1} and~\ref{eq:EffectiveEnergy} retain the same form after the renormalization group steps. We then write the ordinary differential equations for the parameters w.r.t. $s$, called $\beta$ functions~\cite{Amit,Kleinert}. To obtain the contribution to the renormalization of $B_{ijkl}$ from the anharmonic terms, we find it easier to work with the effective free energy $F_\text{eff}$ where the in-plane phonon are completely integrated out (see Fig.~\ref{fig:figure2}). However, for $C_{ijkl}$, it is much simpler to use the free energy $F_\text{eff,1}$ (see Fig.~\ref{fig:figure2_b}). To one-loop order we get:
\begin{figure}[h!]
  \centering
  \includegraphics{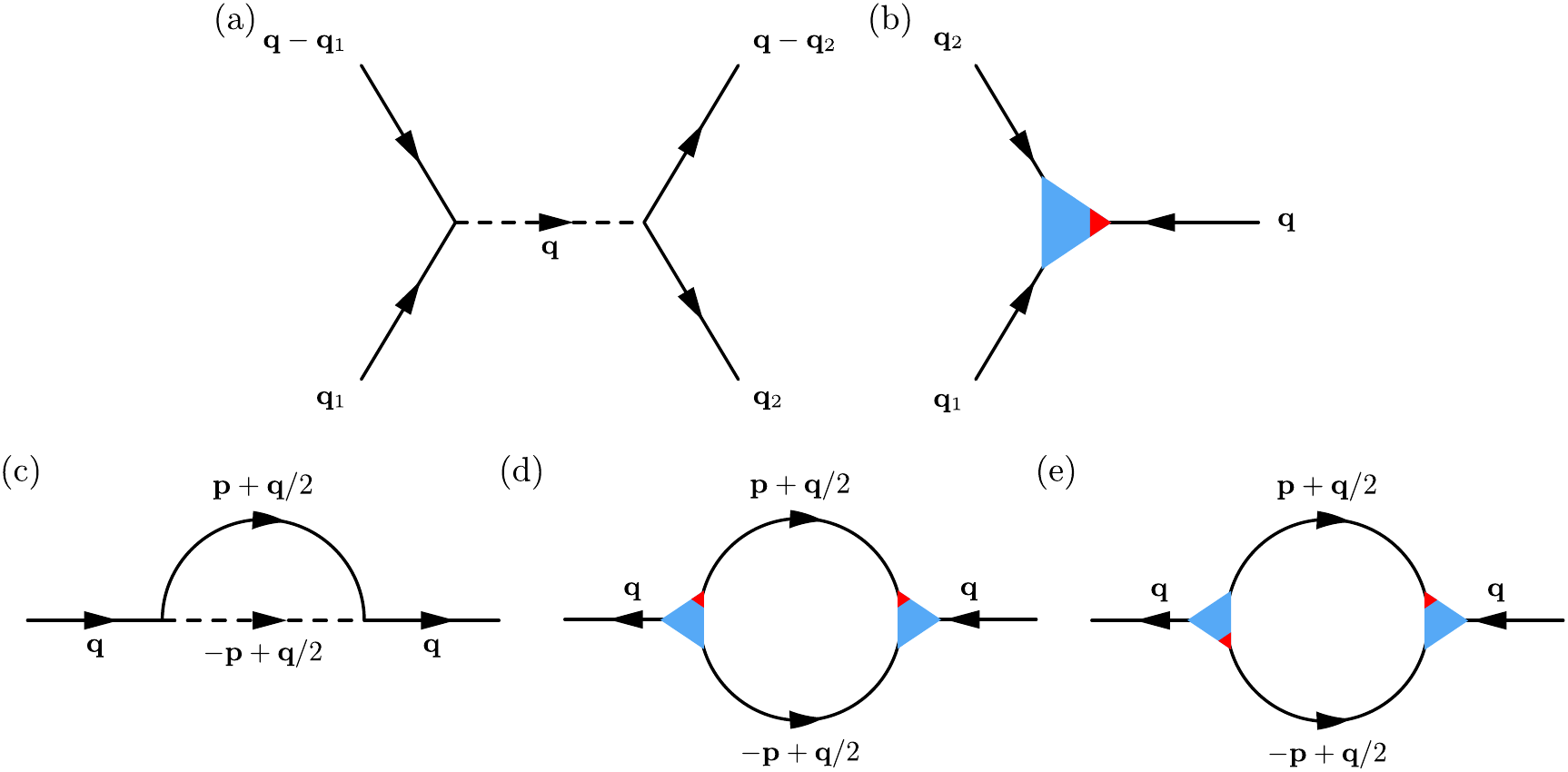}
  \caption{Feynman diagrams relevant to the renormalization of bending rigidity. (a)~Four-point and (b)~three-point vertices describe the quartic and cubic terms in the effective free energy of Eq.~\ref{eq:EffectiveEnergy}, respectively. the straight legs represent radial displacement fields $\tilde{h}(\mathbf{q})$. The red part of the three-point vertex in (b) connects to the field $\tilde{h}(\mathbf{q})$ corresponding to wave vector which is in the argument of $P_{ij}^T$. (c–e)~One-loop diagrams that contribute to the renormalization flows of the bending rigidities $B_{ijkl}^R$. The momentum $\mbf{p}$ in (c), (d), (e) is the loop momentum which is integrated over a shell $\Lambda/b < p < \Lambda$ and whole Fourier space in momentum-shell renormalization and self consistent calculation respectively.}
  \label{fig:figure2}
\end{figure}
\begin{figure}[h!]
  \centering
  \includegraphics{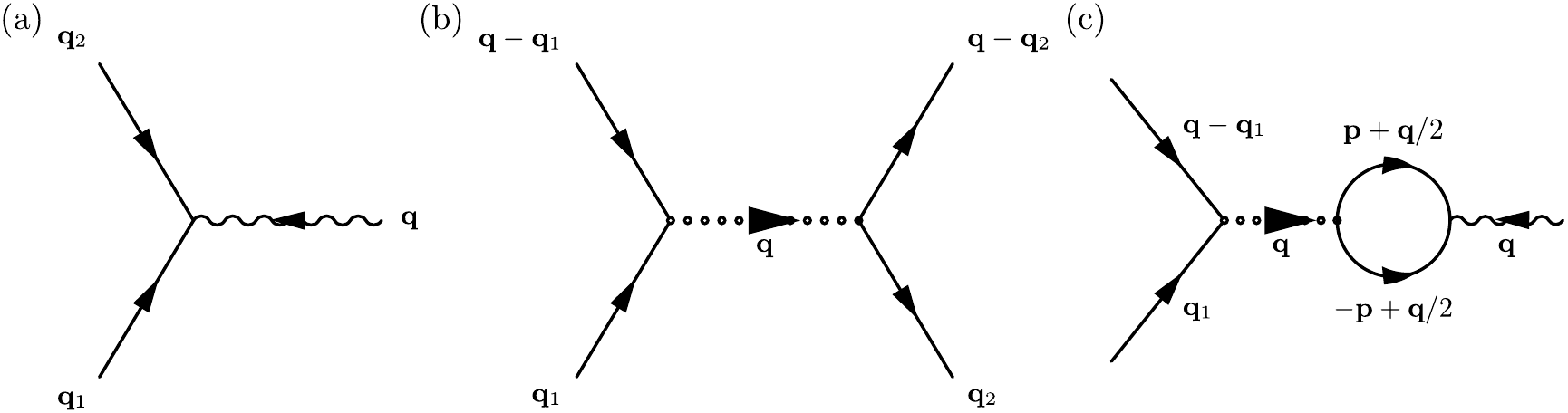}
  \caption{Feynman diagrams relevant to the renormalization of in-plane moduli. (a)~Three-point (b)~four-point vertices describe the cubic $\tilde{u}\tilde{h}\tilde{h}$ and quartic $\tilde{h}\tilde{h}\tilde{h}\tilde{h}$ terms in the free energy of Eq.~\ref{eq:EffectiveEnergy1}, respectively. The wiggly line in (a) is the leg correspoding to in-plane displacement field $\tilde{u}_i(\mbf{q})$. (c)~One-loop diagram that contributes to the renormalization flows of the in-plane stiffnesses $C_{ijkl}^R$ associated with the three-point vertex in (a). The connected legs in these diagrams represent the propagators $G_{hh}(\mathbf{q})$. The momentum $\mbf{p}$ in (c) is the loop momentum which is integrated over a shell $\Lambda/b < p < \Lambda$ and whole Fourier space in momentum-shell renormalization and self consistent calculation respectively.}
  \label{fig:figure2_b}
\end{figure}
\begin{subequations}
\label{eq:BetaFunctions}
\begin{equation}
\label{eq:BetaBending}
\begin{split}
    \beta_{B_{ijkl}} &= \frac{d B'_{ijkl}}{d s} = 2(\zeta_h-1)B'_{ijkl} + \frac{A}{4\pi^2} \frac{d}{ds} \int_{\Lambda/b < |\mbf{p}| < \Lambda} d^2\mbf{p}\, \varepsilon_{im}\varepsilon_{jn}\varepsilon_{kr}\varepsilon_{lt} p_m p_n p_r p_t \frac{N(C_{ijkl}')}{D(C_{ijkl}';\mbf{p})}G_{hh}(\mbf{p})\\
    &\phantom{= \frac{d B'_{\varphi\varphi\varphi\varphi}}{d s} = 2(\zeta_h)} - \frac{2A^2}{4\pi^2 R'^2 k_B T} \frac{d}{ds} \int_{\Lambda/b < |\mbf{p}| < \Lambda} d^2\mbf{p}\, \varepsilon_{im}\varepsilon_{jn}\varepsilon_{kr}\varepsilon_{lt} p_m p_n p_r p_t p_z^4 \frac{N^2(C_{ijkl}')}{D^2(C_{ijkl}';\mbf{p})}G_{hh}^2(\mbf{p}),
\end{split}
\end{equation}
\begin{equation}
\begin{split}
    \beta_{C_{ijkl}} = \frac{d C'_{ijkl}}{d s} = 2(2\zeta_h-1)C'_{ijkl} - \frac{A^2}{8\pi^2 k_B T}C'_{ijmn}C'_{rtkl} \frac{d}{ds} \int_{\Lambda/b < |\mbf{p}| < \Lambda} d^2\mbf{p}\, p_m p_n p_r p_t G_{hh}^2(\mbf{p}),
\end{split}
\end{equation}
\begin{equation}
\begin{split}
    \beta_{\sigma_{ij}} = \frac{d \sigma'_{ij}}{d s} = 2\zeta_h\sigma'_{ij},
\end{split}
\end{equation}
\begin{equation}
\begin{split}
    \beta_{R} = \frac{d R'}{d s} = -R',
\end{split}
\end{equation}
\end{subequations}
where $\varepsilon_{ij}$ is the permutation symbol ($\varepsilon_{11} =\varepsilon_{22} = 0$ and $\varepsilon_{12} = -\varepsilon_{21} = 1$). The scale-dependent parameters $B_{ijkl}'$, $C_{ijkl}'$, and $\sigma_{ij}'$, obtained by integrating the differential equations in Eqs.~\ref{eq:BetaFunctions} up to a scale $s = \ln(\ell^*/a)$ with initial conditions $B_{ijkl}'(0) = \kappa_0 \delta_{ij}\delta_{kl}$, $C_{ijkl}(0) = \lambda_0\delta_{ij}\delta_{kl} +\mu_0(\delta_{ik}\delta_{jl}+\delta_{il}\delta_{jk})$ and $\sigma_{ij}(0) = 0$, are related to the scale-dependent renormalized parameters as $B_{ijkl}^R(s) = B_{ijkl}'(s)e^{(2-2\zeta_h)s}=B_{ijkl}'(s)$, $C_{ijkl}^R(s) = C_{ijkl}'(s)e^{(2-4\zeta_h)s} = C_{ijkl}'(s)e^{-2s}$ and $\sigma_{ij}^R(s) = \sigma_{ij}'(s)e^{-2 s}$, since $\zeta_h = 1$.

With the initial conditions just mentioned, the only term that can make the RG flow anisotropic is the last term in Eq.~\ref{eq:BetaBending}. In the limit $\ell_\text{th} < \ell^* \ll \ell_\text{el}^0$, the second and third terms in Eq.~\ref{eq:BetaBending} are $ \sim k_B T Y'/(\kappa' \Lambda^2)$ and $\sim k_B T Y'^2/(\kappa'^2 R'^2 \Lambda^6)$ respectively. The third term is of the order of the second terms when:
\begin{equation}
\label{eq:LElasticIntuition}
\begin{split}
    k_B T Y'/(\kappa' \Lambda^2) &\lesssim k_B T Y'^2/(\kappa'^2 R'^2 \Lambda^6)\\
    \Rightarrow k_B T Y_R(s)e^{2s}a^2/\kappa_R(s) &\lesssim k_B T Y_R^2(s)e^{4s} a^6/(\kappa^2_R(s) R^2e^{-2s})\\
    \Rightarrow 1 &\lesssim Y_R(s)(\ell/a)^4 a^4/(\kappa_R(s) R^2)\\
    \Rightarrow \ell &\gtrsim (\kappa_R(s) R^2/Y_R(s))^{1/4}
\end{split}
\end{equation}
Comparing the right hand side of the last line of Eq.~\ref{eq:LElasticIntuition} with Eq.~\ref{eq:ElasticLengthHarmonic}, we define a new elastic length scale as
\begin{equation}
\label{eq:ElasticLength}
    \ell_\text{el} \equiv \frac{\pi}{q_\text{el}} = \pi\left(\frac{\kappa_R(\ell_\text{el}) R^2}{Y_R(\ell_\text{el})}\right)^\frac{1}{4},
\end{equation}
where $\kappa_R(\ell_\text{el})$ and $Y_R(\ell_\text{el})$ are themselves dependent on the elastic length scale. We will give a better estimate of elastic length later. From, Eq.~\ref{eq:LElasticIntuition}, we see that the third term in Eq.~\ref{eq:BetaBending}, which is anisotropic, is negligible compared to the isotropic second term in the regime $\ell_\text{th} < \ell^* \ll \ell_\text{el}$, and can be ignored. Therefore, in this region the material parameters remain isotropic. In this limit, the $\beta$-functions are exactly the same as those for a thermally fluctuating isotropic sheet~\cite{AndrejRibbon}. Hence, in this regime, the scale dependence of the isotropic material parameters is the same as in case of isotropic sheets:
\begin{equation}
\begin{split}
    \frac{\kappa_R(\ell)}{\kappa_0} &\approx (\ell/\ell_\text{th})^\eta, \ell_\text{th} \ll \ell \ll \ell_\text{el},\\
    \frac{\lambda_R(\ell)}{Y_0},\frac{\mu_R(\ell)}{Y_0},\frac{Y_R(\ell)}{Y_0} &\approx c(\ell/\ell_\text{th})^{-\eta_u}, \ell_\text{th} \ll \ell \ll \ell_\text{el},
\end{split}
\end{equation}
where the prefactor $c$ is $-0.36$, $0.72$ and $1.0$ for $\lambda_R$, $\mu_R$ and $Y_R$ respectively~\cite{AndrejRibbon}, $\eta \approx 0.8 - 0.85$~\cite{AndrejRibbon,radzihovskySCSA,Mouhanna2009}, and the exponents $\eta$ and $\eta_u$ satisfy the identity $2\eta+\eta_u = 2$, which is a result of infinitesimal rotational symmetry of the system about in-plane axes~\cite{guitter1989thermodynamical}. Replacing These expressions of $\kappa_R(\ell)$ and $Y_R(\ell)$ in Eq.~\ref{eq:ElasticLength} we get the following estimate of elastic length:
\begin{equation}
\label{eq:ElasticLengthExp}
\ell_\text{el} = \frac{\pi}{q_\text{el}} = \left(\frac{\kappa_0 R^2 q_\text{th}^{\eta+\eta_u}}{Y_0}\right)^\frac{1}{4-\eta-\eta_u}.
\end{equation}
A better estimate of the thermal length scale can also be obtained from Eq.~\ref{eq:BetaBending}. This is the scale, $\ell = a e^s$, at which the second term is comparable to the first (Gaussian) term of Eq.~\ref{eq:BetaBending} is the system size where the anharmonicity of the free energy becomes important. Then, by simplifying the second term in Eq.~\ref{eq:BetaBending} using the fact that the material parameters are isotropic and using the expressions $\kappa'(s) = \kappa_R(s) \approx \kappa_0$ (the last equality is due to the fact that the parameters only start to renormalize at the thermal length scale), $Y'(s) = Y_R(s)e^{2s} \approx Y_0e^{2s} \approx \kappa_0(s)$, $s = \ln(\ell_\text{th}/a)$, $\Lambda = \pi/a$, we get~\cite{AndrejRibbon}:
\begin{equation}
    \ell_\text{th} \equiv \frac{\pi}{q_\text{th}} =\sqrt{\frac{16\pi^3 \kappa_0^2}{3k_BTY_0}}.
\end{equation}
\end{itemize}
\subsection{Scaling analysis for $\ell^* \gg \ell_\text{el}$}
\label{sec:AnisotropicScaling}
\begin{figure}[b]
  \centering
  \includegraphics{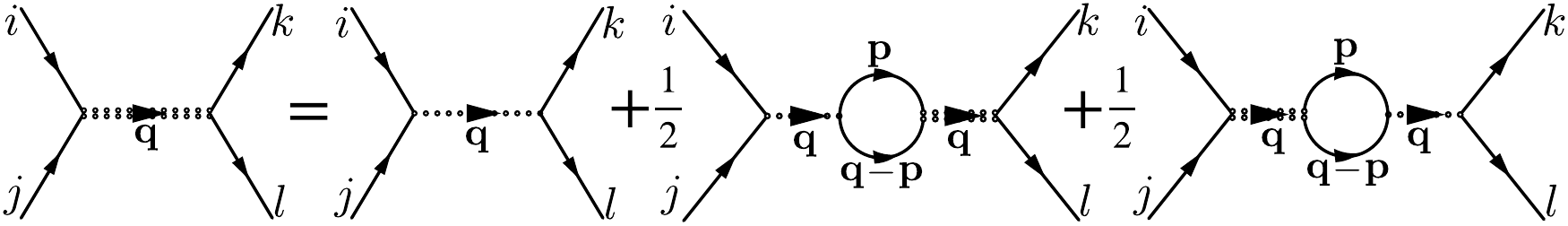}
  \caption{The self consistent perturbative equation for $C_{ijkl}^R(\mathbf{q})$ is shown diagrammatically using the $C_{ijkl}\partial_i\tilde{h}\partial_j\tilde{h}\partial_k\tilde{h}\partial_l\tilde{h}$ vertex. The symmetrization is due to the major symmetry of the Hamiltonian. The dotted line indicates $C_{ijkl}^R(q_{\text{el}})$ and the doubled dotted line $C_{ijkl}^R(\mathbf{q})$.}
  \label{fig:figureSCSA}
\end{figure}

Thus far, in our discussion, we have seen that up to the elastic length scale the parameters remain isotropic and $B_{ijkl}^R(q_\text{el}) \approx \kappa_R(q_\text{el})\delta_{ij}\delta_{kl} \approx \delta_{ij}\delta_{kl}\kappa_0 (q_\text{el}/q_\text{th})^{-\eta}$ and $C_{ijkl}^R(q_\text{el}) \approx \lambda_R(q_\text{el})\delta_{ij}\delta_{kl}+\mu_R(q_\text{el})(\delta_{ik}\delta_{jl}+\delta_{il}\delta_{jk}) \approx Y_0(-0.36\delta_{ij}\delta_{kl}+0.72(\delta_{ik}\delta_{jl}+\delta_{il}\delta_{jk})) (q_\text{el}/q_\text{th})^{\eta_u}$. Therefore, integrating out the degrees of freedom on scales smaller than $\ell_\text{el}$, the free energy takes the form:
\begin{subequations}
\label{eq:EffectiveEnergyCourseGrained}
\begin{equation}
\begin{split}
    F^0_\text{eff}(q_\text{el})/A &= \sum_{\substack{\mathbf{q} \neq 0\\ q < q_\text{el}}}\frac{1}{2}\left(B_{ijkl}^R (q_\text{el}) q_i q_j q_k q_l + \frac{q_z^4 N(C_{ijkl}^R(q_\text{el}))}{R^2 D(C_{ijkl}^R(q_\text{el});\mbf{q})}\right)\tilde{h}(\mathbf{q})\tilde{h}(-\mathbf{q})\\
    &\approx \sum_{\substack{\mathbf{q} \neq 0\\ q < q_\text{el}}}\frac{1}{2}\left(\kappa_R (q_\text{el}) q^4 + \frac{q_z^4 Y_R(q_\text{el})}{R^2 q^4}\right)\tilde{h}(\mathbf{q})\tilde{h}(-\mathbf{q}),
\end{split}
\end{equation}
\begin{equation}
\begin{split}
    F^I_\text{eff}(q_\text{el})/A &= \sum_{\substack{\mathbf{q}_1+\mathbf{q}_2=-\mathbf{q} \neq 0\\q_1,q_2,q<q_\text{el}}} \frac{q_z^2}{2q^2}[q_{1i}P^T_{ij}(\mathbf{q})q_{2j}]\frac{N(C_{ijkl}^R(q_\text{el})) q^4}{RD(C_{ijkl}^R(q_\text{el});\mathbf{q})}\tilde{h}(\mathbf{q})\tilde{h}(\mathbf{q}_1)\tilde{h}(\mathbf{q}_2)\\
    &\phantom{=} + \sum_{\substack{\mathbf{q}_1+\mathbf{q}_2=\mathbf{q} \neq 0\\ \mathbf{q}_3+\mathbf{q}_4=-\mathbf{q} \neq 0\\q_1,q_2,q_3,q_4,q<q_\text{el}}} \frac{1}{8}[q_{1i}P^T_{ij}(\mathbf{q})q_{2j}][q_{3i}P^T_{ij}(\mathbf{q})q_{4j}]\frac{N(C_{ijkl}^R(q_\text{el})) q^4}{D(C_{ijkl}^R(q_\text{el});\mathbf{q})}\tilde{h}(\mathbf{q}_1)\tilde{h}(\mathbf{q}_2)\tilde{h}(\mathbf{q}_3)\tilde{h}(\mathbf{q}_4),\\
    &\approx \sum_{\substack{\mathbf{q}_1+\mathbf{q}_2=-\mathbf{q} \neq 0\\q_1,q_2,q<q_\text{el}}} [q_{1i}P^T_{ij}(\mathbf{q})q_{2j}]\frac{Y_R(q_\text{el})q_z^2}{2Rq^2}\tilde{h}(\mathbf{q})\tilde{h}(\mathbf{q}_1)\tilde{h}(\mathbf{q}_2)\\
    &\phantom{\approx} + \sum_{\substack{\mathbf{q}_1+\mathbf{q}_2=\mathbf{q} \neq 0\\ \mathbf{q}_3+\mathbf{q}_4=-\mathbf{q} \neq 0\\q_1,q_2,q_3,q_4,q<q_\text{el}}} \frac{Y_R(q_\text{el})}{8}[q_{1i}P^T_{ij}(\mathbf{q})q_{2j}][q_{3i}P^T_{ij}(\mathbf{q})q_{4j}]\tilde{h}(\mathbf{q}_1)\tilde{h}(\mathbf{q}_2)\tilde{h}(\mathbf{q}_3)\tilde{h}(\mathbf{q}_4).
\end{split}
\end{equation}
\end{subequations}
Starting from this course-grained free energy, the harmonic approximation to the Green's function is:
\begin{equation}
    G_{hh}^0(\mathbf{q};q<q_\text{el}) = \frac{k_B T/A}{\kappa_R(q_\text{el}) q^4 + \frac{Y_R(q_\text{el}) q_z^4}{R^2 q^4}},
\end{equation}
which is aniostropic for $q<q_\text{el}$. Now, following the argument of \cite{RadzihovskyToner} (section V), the regime of wavevectors that dominates the $\tilde{h}$-fluctuations is $q^8 \approx (q_\text{el} q_z)^4$ i.e., $q_z = q_{\varphi}^2/q_\text{el}$. Therefore, for small $q \ll q_\text{el}$, $q_z \sim q_{\varphi}^{2} \ll q_{\varphi}$. This leads to a simplification of the expression of $D(C_{ijkl};\mbf{q})$ keeping only the lowest order terms in $q_\varphi$:
\begin{equation}
    \frac{q_z^4 N(C_{ijkl}^R(q_\text{el}))}{D(C_{ijkl}^R(q_\text{el});\mbf{q})} \approx \frac{C_{\varphi \varphi \varphi \varphi}^R(q_\text{el}) C_{\varphi z \varphi z}^R(q_\text{el}) C_{z z z z}^R(q_\text{el})-C_{\varphi \varphi z z}^R(q_\text{el})^2 C_{\varphi z \varphi z}^R(q_\text{el})}{C_{\varphi \varphi \varphi \varphi}^R(q_\text{el}) C_{\varphi z \varphi z}^R(q_\text{el})}\frac{q_z^4}{q_{\varphi}^4}=Y_{zz}^R(q_\text{el})\frac{q_z^4}{q_{\varphi}^4}.
\end{equation}
In this regime, if we count the dimension of wave vector component $q_\varphi$ as $[q_\varphi] = 1$, we have to count the dimension of $q_z$ as $[q_z] = 2$ since $q_z \sim q_\varphi^2$. This happens in strongly anisotropic systems, see for example~\cite{DiehlShpot}. In fact the scaling theory that will be presented in this paper is identical to that of tubules found in \cite{RadzihovskyToner} as well as that of membranes under uni-axially tension \cite{bahri2022mechanical}. Though these are different systems physically, the scaling of the correlation functions and the arguments to obtain these scalings are identical. Then, the dimension of area $A$ is $[A] = -1 - 2 = -3$. With these, and requiring that Gaussian part of the effective free energy
\begin{equation}
    F^0_\text{eff}(q_\text{el}) \approx \sum_{\substack{\mathbf{q} \neq 0\\ q < q_\text{el}}}\frac{A}{2}\left(B_{ijkl}^R (q_\text{el}) q_i q_j q_k q_l + \frac{q_z^4 Y^R_{zz}(q_\text{el})}{R^2 q_\varphi^4}\right)\tilde{h}(\mathbf{q})\tilde{h}(-\mathbf{q})
\end{equation}
dimensionless, we get the following na\"ve dimensions:
\begin{equation}
\label{eq:dimension1}
    [\tilde{h}(\mathbf{q})] = -1/2, [B_{\varphi\varphi\varphi\varphi}^R(q_\text{el})] = 0, [B_{\varphi\varphi zz}^R(q_\text{el})] = -2, [B_{zzzz}^R(q_\text{el})] = -4, [Y_{zz}^R(q_\text{el})/R^2] = 0
\end{equation}
Therefore, the terms $B_{\varphi\varphi zz}^R(q_\text{el})q_\varphi^2 q_z^2 \tilde{h}(\mathbf{q})\tilde{h}(-\mathbf{q})$, $B_{zzzz}^R(q_\text{el})q_z^4 \tilde{h}(\mathbf{q})\tilde{h}(-\mathbf{q})$ are irrelevant. Keeping only the relevant terms then the harmonic part of the free energy is:
\begin{equation}
    F^0_\text{eff}(q_\text{el}) \approx \sum_{\substack{\mathbf{q} \neq 0\\ q < q_\text{el}}}\frac{A}{2}\left(B_{\varphi\varphi\varphi\varphi}^R (q_\text{el}) q_\varphi^4 + \frac{q_z^4 Y^R_{zz}(q_\text{el})}{R^2 q_\varphi^4}\right)\tilde{h}(\mathbf{q})\tilde{h}(-\mathbf{q}),
\end{equation}
and the harmonic Green's function can be approximated as:
\begin{equation}
    G_{hh}^0(\mathbf{q};q<q_\text{el}) \approx \frac{k_B T/A}{B_{\varphi\varphi\varphi\varphi}^R (q_\text{el}) q_\varphi^4 + \frac{Y^R_{zz}(q_\text{el}) q_z^4}{R^2 q_\varphi^4}}.
\end{equation}
To get the na\"ive dimensions of other moduli, first go back to the form of the in-plane strain tensor and require that all the terms in the same component of strain have the same dimension. This way we get:
\begin{equation}
\label{eq:dimension2}
\begin{split}
    \tilde{u}_{\varphi\varphi}&: [\partial_\varphi \tilde{u}_\varphi] = [\tilde{h}/R] = [(\partial_\varphi \tilde{h})^2] = 1 \Rightarrow [\tilde{u}_\varphi] = 0, [1/R] = 3/2,\\
    \tilde{u}_{zz}&: [\partial_z \tilde{u}_z] = [(\partial_z \tilde{h})^2] = 3 \Rightarrow [\tilde{u}_z] = 1,\\
    \tilde{u}_{\varphi z}&: [\partial_\varphi \tilde{u}_z] = [\partial_z \tilde{u}_\varphi] = [(\partial_z \tilde{h})(\partial_\varphi \tilde{h})] = 2,
\end{split}
\end{equation}
where we used $\tilde{u}_{\varphi\varphi}$ and $\tilde{u}_{zz}$ get the dimensions of $\tilde{u}_\varphi$, $\tilde{u}_z$ and $1/R$, and checked their consistency with $\tilde{u}_{\varphi z}$. With these dimensions, we use the free energy in Eq.~\ref{eq:EffectiveEnergy1} to find the dimensions of $C_{ijkl}^R(q_\text{el})$:
\begin{equation}
    [C_{\varphi\varphi\varphi\varphi}^R(q_\text{el})] = 1, [C_{\varphi\varphi zz}^R(q_\text{el})] = [C_{\varphi z \varphi z}^R(q_\text{el})] = -1, [C_{z z z z}^R(q_\text{el})] = -3.
\end{equation}
This means that only terms with $C_{\varphi\varphi\varphi\varphi}^R(q_\text{el})$ as coefficient are relevant. Keeping only terms with coefficients $C_{\varphi\varphi\varphi\varphi}^R(q_\text{el})$ in Eq.~\ref{eq:EffectiveEnergy1}, if we integrate out the in-plane displacements, the interacting part $F_\text{eff}^I(q_\text{el})$ of effective free energy is $F_\text{eff}^I(q_\text{el}) = 0$. Therefore, the effective free energy:
\begin{equation}
    F_\text{eff}(q_\text{el}) \approx F^0_\text{eff}(q_\text{el}) \approx \sum_{\substack{\mathbf{q} \neq 0\\ q < q_\text{el}}}\frac{A}{2}\left(B_{\varphi\varphi\varphi\varphi}^R (q_\text{el}) q_\varphi^4 + \frac{q_z^4 Y^R_{zz}(q_\text{el})}{R^2 q_\varphi^4}\right)\tilde{h}(\mathbf{q})\tilde{h}(-\mathbf{q})
\end{equation}
describes a free field theory and $B_{\varphi\varphi\varphi\varphi}^R(q_\text{el})$ and $Y_{zz}^R(q_\text{el})$ do not renormalize any further as we integrate out Fourier modes beyond $q <q_\text{el}$:
\begin{subequations}
\begin{equation}
    B_{\varphi\varphi\varphi\varphi}^R(\mathbf{q}) \approx B_{\varphi\varphi\varphi\varphi}^R(q_\text{el}) = \kappa_R(q_\text{el}) \approx \kappa_0 (q_\text{el}/q_\text{th})^{-\eta},
\end{equation}
\begin{equation}
\label{eq:Yzz}
    Y_{zz}^R(\mathbf{q}) \approx Y_{zz}^R(q_\text{el}) = Y_R(q_\text{el}) \approx Y_0 (q_\text{el}/q_\text{th})^{\eta_u}.
\end{equation}
\end{subequations}
Hence, the Green's function $G_{hh}(\mathbf{q};q<q_\text{el}) = G_{hh}^0(\mathbf{q};q<q_\text{el})$. However, the free energy in Eq.~\ref{eq:EffectiveEnergy1} is not a free field theory because $C_{\varphi\varphi\varphi\varphi}^R(q_\text{el})$ is relevant. Then we can use the Feynman diagram in Fig.~\ref{fig:figureSCSA} to write a self consistent perturbative equation (see a similar analysis in~\cite{RadzihovskyToner}, see also~\cite{bahri2022mechanical} for more details):
\begin{equation}
\label{eq:C1111Recursion}
\begin{split}
    C_{\varphi\varphi\varphi\varphi}^R(\mathbf{q}) &=  C_{\varphi\varphi\varphi\varphi}^R(q_\text{el}) - \frac{A^2}{8\pi^2 k_B T}(C^R_{\varphi\varphi\varphi\varphi}(\mathbf{q})C^R_{\varphi\varphi\varphi\varphi}(q_\text{el})) \int_{|\mbf{p}| < q_\text{el}} dp_\varphi dp_z \, p_\varphi^2 (p_\varphi -q_\varphi)^2 G_{hh}(\mbf{p})G_{hh}(\mbf{p}-\mbf{q})\\
    &= C_{\varphi\varphi\varphi\varphi}^R(q_\text{el})-\frac{k_B T}{8\pi^2}(C^R_{\varphi\varphi\varphi\varphi}(\mathbf{q})C^R_{\varphi\varphi\varphi\varphi}(q_\text{el})) \int_{|\mbf{p}| < q_\text{el}} dp_\varphi dp_z \, p_\varphi^2 (p_\varphi -q_\varphi)^2 \times\\
    &\hspace{10mm}\frac{1}{\left(B_{\varphi\varphi\varphi\varphi}^R (q_\text{el}) p_\varphi^4 + \frac{Y^R_{zz}(q_\text{el}) p_z^4}{R^2 p_\varphi^4}\right)\left(B_{\varphi\varphi\varphi\varphi}^R (q_\text{el}) (p_\varphi-q_\varphi)^4 + \frac{Y^R_{zz}(q_\text{el}) (p_z-q_z)^4}{R^2 (p_\varphi-q_\varphi)^4}\right)}.
\end{split}
\end{equation}
One can extract how the integral scales with $q_\varphi$ or $q_z$ by non dimensionalizing $p_\varphi$ and $p_z$ with either $q_\varphi$ and $q_\varphi^2$ or $q_z^{1/2}$ and $q_z$ respectively. By doing this one can find that the integral scales as $q_\varphi^{-1} \sim q_z^{-1/2}$. This observation tells us that the self consistent equation can be solved by the ansatz:
\begin{equation}
\label{eq:ScaleCphi}
    C_{\varphi\varphi\varphi\varphi}^R(\mathbf{q}) = q_\varphi^{\eta_\varphi} \Omega_{\varphi\varphi\varphi\varphi}(q_\varphi/(q_\text{el}q_z)^z).
\end{equation}
Using simple power counting, we find that
\begin{equation}
    \eta_\varphi = 1, z = 1/2.
\end{equation}
From the form of the integral in Eq.~\ref{eq:C1111Recursion}, it is easy to see that the function $\Omega_{\varphi\varphi\varphi\varphi}$ is independent of $q_\varphi$ when $q_\varphi \rightarrow 0$, as well as independent of $q_z$ when $q_z \rightarrow 0$. This means the following:
\begin{equation}
    \Omega_{\varphi\varphi\varphi\varphi}(x) \propto \begin{cases} \text{consant}, &x\rightarrow \infty\\ x^{-\eta_\varphi}, & x \rightarrow 0\end{cases}.
\end{equation}
This implies the following:
\begin{equation}
    C_{\varphi\varphi\varphi\varphi}^R(\mathbf{q}) \propto \begin{cases} q_\varphi, & q < q_\text{el} < q_\text{th} \text{ and } q_\varphi \gg (q_\text{el}q_z)^{1/2}\\\sqrt{q_z}, & q < q_\text{el} < q_\text{th} \text{ and } q_\varphi \ll (q_\text{el}q_z)^{1/2} \end{cases}.
\end{equation}
Although the other moduli are irrelevant, we can repeat this same analysis to obtain how they scale. We can check how these moduli scale in our simulations and therefore better verify our theory and provide an understanding of the mechanical properties of nanotubes. We can check for example how the shear modulus should scale:
\begin{equation}
\begin{split}
   C_{\varphi z \varphi z}^R(\mbf{q})  &= C_{\varphi z \varphi z}^R(q_\text{el}) -  \frac{k_BT}{2\pi^2}C_{\varphi z \varphi z}^R(\mbf{q})C_{\varphi z \varphi z}^R(q_\text{el}) \int dp_\varphi  dp_z  (p_\varphi-q_\varphi)(p_z-q_z) p_\varphi p_z \times\\
   &\hspace{10mm}\frac{1}{\left(B_{\varphi\varphi\varphi\varphi}^R (q_\text{el}) p_\varphi^4 + \frac{Y^R_{zz}(q_\text{el}) p_z^4}{R^2 p_\varphi^4}\right)\left(B_{\varphi\varphi\varphi\varphi}^R (q_\text{el}) (p_\varphi-q_\varphi)^4 + \frac{Y^R_{zz}(q_\text{el}) (p_z-q_z)^4}{R^2 (p_\varphi-q_\varphi)^4}\right)}.
\end{split}
\end{equation}
By non dimensionalizing $p_\varphi$ and $p_z$ with either $q_\varphi$ and $q_\varphi^2$ respectively, we find the integral scales as $q_\varphi \sim q_z^{1/2} \rightarrow 0$ as $q\rightarrow 0$. This implies that $C_{\varphi z \varphi z}^R(\mbf{q}) \approx C_{\varphi z \varphi z}^R(q_\text{el})$:
\begin{equation}
    C_{\varphi z \varphi z}^R(\mbf{q}) \approx C_{\varphi z \varphi z}^R(q_\text{el}), \text{ }q < q_\text{el} < q_\text{th}.
\end{equation}
Similarly, for $C_{\varphi \varphi zz}^R(\mbf{q})$, we have the following self-consistent equation:
\begin{equation}
\begin{split}
   C_{\varphi\varphi zz}^R(\mbf{q})  = & C_{\varphi\varphi zz}^R(q_\text{el})\\ & -  \frac{k_BT}{16\pi^2}\int dp_\varphi  dp_z \frac{[C_{\varphi\varphi\varphi\varphi}^R(\mbf{q}) (p_\varphi-q_\varphi)^2+C_{\varphi\varphi zz}^R(\mbf{q}) (p_z-q_z)^2] [C_{\varphi\varphi z z}^R(q_\text{el}) p_\varphi^2+C_{zzzz}^R(q_\text{el}) p_z^2] }{\left(B_{\varphi\varphi\varphi\varphi}^R (q_\text{el}) p_\varphi^4 + \frac{Y^R_{zz}(q_\text{el}) p_z^4}{R^2 p_\varphi^4}\right)\left(B_{\varphi\varphi\varphi\varphi}^R (q_\text{el}) (p_\varphi-q_\varphi)^4 + \frac{Y^R_{zz}(q_\text{el}) (p_z-q_z)^4}{R^2 (p_\varphi-q_\varphi)^4}\right)} \\ &
   -  \frac{k_BT}{16\pi^2}\int dp_\varphi  dp_z \frac{[C_{\varphi\varphi\varphi\varphi}^R(q_\text{el}) (p_\varphi-q_\varphi)^2+C_{\varphi\varphi zz}^R(q_\text{el}) (p_z-q_z)^2] [C_{\varphi\varphi z z}^R(\mbf{q}) p_\varphi^2+C_{zzzz}^R(\mbf{q}) p_z^2] }{\left(B_{\varphi\varphi\varphi\varphi}^R (q_\text{el}) p_\varphi^4 + \frac{Y^R_{zz}(q_\text{el}) p_z^4}{R^2 p_\varphi^4}\right)\left(B_{\varphi\varphi\varphi\varphi}^R (q_\text{el}) (p_\varphi-q_\varphi)^4 + \frac{Y^R_{zz}(q_\text{el}) (p_z-q_z)^4}{R^2 (p_\varphi-q_\varphi)^4}\right)}.
\end{split}
\end{equation}
Since the $C_{zzzz}^R(\mathbf{q})$ term is less relevant (has a lower scaling dimension) than the $C_{\varphi \varphi zz}^R(\mathbf{q})$ term, we may ignore its contribution to the equation. In addition $C_{zzzz}^R(q_\text{el})p_z^2$ scales with a higher power of $q_\varphi$ than $C_{\varphi \varphi zz}(q_\text{el})p_\varphi^2$ and may thus be ignored. The integral with coefficient $C_{\varphi\varphi\varphi\varphi}^R(\mbf{q}) C_{\varphi\varphi zz}^R(q_\text{el})$ scales as $q_\varphi^{-1}$ and the integral with coefficient $C_{\varphi\varphi zz}^R(\mbf{q}) C_{\varphi\varphi zz}^R(q_\text{el})$ scales as $q_\varphi$. Finally, the integral with coefficient $C_{\varphi\varphi\varphi\varphi}(q_\text{el}) C_{\varphi\varphi zz}^R(\mbf{q})$ scales as $q_\varphi^{-1}$. In addition, as we have seen before, $C_{\varphi\varphi\varphi\varphi}^R(\mbf{q}) \sim q_\varphi$. This means that the self consistent equation gives the following:
\begin{equation}
\begin{split}
    &C_{\varphi\varphi z z}^R(\mbf{q}) \approx \text{const.}\times C_{\varphi\varphi z z}^R(q_\text{el}) + (\text{const.}\times q_\varphi + \text{const.}\times q_\varphi^{-1})\times C_{\varphi\varphi z z}^R(\mbf{q}), \text{ }q < q_\text{el} < q_\text{th},\\
    \Rightarrow & C_{\varphi\varphi z z}^R(\mbf{q}) (1+(\text{const.}\times q_\varphi + \text{const.}\times q_\varphi^{-1})) = \text{const.}\times C_{\varphi\varphi z z}^R(q_\text{el}), \text{ }q < q_\text{el} < q_\text{th}.
\end{split}
\end{equation}
Hence, for the two sides in the above equation to have the same scaling as $q_\varphi \rightarrow 0$, it is necessary that $C_{\varphi \varphi zz}^R(\mathbf{q}) \sim q_{\varphi}$ and thus more explicitly:
\begin{equation}
    C_{\varphi\varphi zz}^R(\mathbf{q}) \propto \begin{cases} q_\varphi, & q < q_\text{el} < q_\text{th} \text{ and } q_\varphi \gg (q_\text{el}q_z)^{1/2}\\\sqrt{q_z}, & q < q_\text{el} < q_\text{th} \text{ and } q_\varphi \ll (q_\text{el}q_z)^{1/2} \end{cases}.
\end{equation}

For $C_{zzzz}^R(\mbf{q})$, we have the following self-consistent equation:
\begin{equation}
\label{eq:C2222Recursion}
\begin{split}
   C_{zzzz}^R(\mbf{q})  &= C_{zzzz}^R(q_\text{el}) -  \frac{k_BT}{8\pi^2} \int dp_\varphi  dp_z \\ &  [C^R_{zzzz}(\mathbf{q})(p_z-q_z)q_z + C^R_{\varphi \varphi zz}(\mathbf{q})(p_{\varphi}-q_{\varphi})q_{\varphi}][C^R_{zzzz}(q_\text{el})(p_z-q_z)q_z + C^R_{\varphi \varphi zz}(q_\text{el})(p_{\varphi}-q_{\varphi})q_{\varphi}]\times\\
   &\hspace{10mm}\frac{1}{\left(B_{\varphi\varphi\varphi\varphi}^R (q_\text{el}) p_\varphi^4 + \frac{Y^R_{zz}(q_\text{el}) p_z^4}{R^2 p_\varphi^4}\right)\left(B_{\varphi\varphi\varphi\varphi}^R (q_\text{el}) (p_\varphi-q_\varphi)^4 + \frac{Y^R_{zz}(q_\text{el}) (p_z-q_z)^4}{R^2 (p_\varphi-q_\varphi)^4}\right)}.
\end{split}
\end{equation}
By non dimensionalizing $p_\varphi$ and $p_z$ with $q_\varphi$ and $q_\varphi^2$ respectively, we find the integral scales as $q_\varphi^3 \sim q_z^{3/2} \rightarrow 0$ as $q_\varphi\rightarrow 0$. This implies that in the limit of $q_\varphi\rightarrow 0$, $C_{z z z z}^R(\mbf{q}) \approx C_{z z z z}^R(q_\text{el})$:
\begin{equation}
    C_{z z z z}^R(\mbf{q}) \approx C_{z z z z}^R(q_\text{el}), \text{ }q < q_\text{el} < q_\text{th}.
\end{equation}
Earlier, we found that $Y_{zz}^R(\mbf{q})$ stops renormalizing in the regime $q < q_\text{el} < q_\text{th}$. But we know that $Y_{zz} = C_{zzzz} -C_{\varphi\varphi zz}^2/C_{\varphi\varphi\varphi\varphi}$. Since $C_{zzzz}^R   \sim \text{constant}$ and $(C_{\varphi\varphi zz}^R)^2/C_{\varphi\varphi\varphi\varphi}^R \sim q_\varphi^2/q_\varphi = q_\varphi \rightarrow 0$ as $q_\varphi \rightarrow 0$, and therefore $C_{zzzz}^R - (C_{\varphi\varphi zz}^R)^2/C_{\varphi\varphi\varphi\varphi}^R \sim \text{constant}$ which matches with our result for $Y_{zz}^R$. Similarly, $Y_{\varphi\varphi} = C_{\varphi\varphi\varphi\varphi} -C_{\varphi\varphi zz}^2/C_{zzzz} \sim q_\varphi$. We summarize these scalings in Table~\ref{tab:Scaling1}.
\begin{table}[h]
\centering
\caption{Scaling functions for $\ell_\text{th}<\ell_\text{el} \Rightarrow q_\text{th}>q_\text{el}$}
\label{tab:Scaling1}
\begin{tabular}{ |c|c|c|c|  }
\hline
Scale &  $q > q_\text{th} > q_\text{el}$ & $q_\text{th} > q >q_\text{el}$& $q_\text{th} > q_\text{el}>q$\\
\hline
$C_{\varphi\varphi\varphi\varphi}^R/Y_0$ &   $\frac{1}{1-\nu_0^2}$  & $\left( \frac{q}{q_\text{th}} \right)^{\eta_u}$  & $\left( \frac{q_\text{el}}{q_\text{th}} \right)^{\eta_u} \left(\frac{q_\varphi}{q_\text{el}}\right)\Omega_{\varphi\varphi\varphi\varphi}\left(q_\varphi/(q_\text{el}q_z)^{1/2}\right) $\\
$C_{\varphi\varphi zz}^R/Y_0$ &   $\frac{\nu_0}{1-\nu_0^2}$  & $\left( \frac{q}{q_\text{th}} \right)^{\eta_u}$  & $\left( \frac{q_\text{el}}{q_\text{th}} \right)^{\eta_u} \left(\frac{q_\varphi}{q_\text{el}}\right)\Omega_{\varphi\varphi zz}\left(q_\varphi/(q_\text{el}q_z)^{1/2}\right) $\\
$C_{\varphi z \varphi z}^R/Y_0$ &   $\frac{1}{2(1+\nu_0)}$  & $\left( \frac{q}{q_\text{th}} \right)^{\eta_u}$  & $\left( \frac{q_\text{el}}{q_\text{th}} \right)^{\eta_u}$\\
$C_{z z z z}^R/Y_0$ &   $\frac{1}{1-\nu_0^2}$  & $\left( \frac{q}{q_\text{th}} \right)^{\eta_u}$  & $\left( \frac{q_\text{el}}{q_\text{th}} \right)^{\eta_u}$\\
$Y_{\varphi\varphi}^R/Y_0$ &   $1$  & $\left( \frac{q}{q_\text{th}} \right)^{\eta_u}$  & $\left( \frac{q_\text{el}}{q_\text{th}} \right)^{\eta_u} \left(\frac{q_\varphi}{q_\text{el}}\right)\Omega_{Y_{\varphi\varphi}}\left(q_\varphi/(q_\text{el}q_z)^{1/2}\right)$\\
$Y_{zz}^R/Y_0$ &   $1$  & $\left( \frac{q}{q_\text{th}} \right)^{\eta_u}$  & $\left( \frac{q_\text{el}}{q_\text{th}} \right)^{\eta_u}$\\
$B_{\varphi\varphi\varphi\varphi}^R/\kappa_0$ &   $1$  & $\left( \frac{q}{q_\text{th}} \right)^{-\eta}$ & $\left( \frac{q_\text{el}}{q_\text{th}} \right)^{-\eta}$\\
 \hline
\end{tabular}
\end{table}

In the beginning of this section, we assumed $\ell_\text{el} \gg \ell_\text{th}$. However, even if $\ell_\text{th} > \ell_{el}$, all the analysis starting from the na\"ive dimensions in Eqs.~\ref{eq:dimension1} and~\ref{eq:dimension2} would remain the same in the regime $\ell > \ell_\text{th} > \ell_{el}$ except the fact that we would our starting course-grained free energy would be $F_\text{eff}(q_\text{th})$ instead of $F_\text{eff}(q_\text{el})$ in Eq.~\ref{eq:EffectiveEnergyCourseGrained} and the material parameters in the course-grained free energy $F_\text{eff}(q_\text{th})$ would be $B_{ijkl}^R(q_\text{th}) \approx B_{ijkl}^0 = \kappa_0 \delta_{ij}\delta_{kl}$ and $C_{ijkl}^R(q_\text{th}) \approx C_{ijkl}^0 = \lambda_0 \delta_{ij}\delta_{kl} +\mu_0(\delta_{ik}\delta_{jl}+\delta_{il}\delta_{jk})$ instead of $B_{ijkl}^R(q_\text{el})$ and $C_{ijkl}^R(q_\text{el})$ in $F_\text{eff}(q_\text{el})$ since the elastic moduli do not renormalize in the regime $q>q_\text{th}$. We summarize these scalings in Table~\ref{tab:Scaling2}.
\begin{table}[h]
\centering
\caption{Scaling Exponents for $\ell_\text{el}< \ell_\text{th} \Rightarrow q_\text{el}> q_\text{th}$}
\label{tab:Scaling2}
\begin{tabular}{ |c|c|c|c|  }
\hline
Scale &  $q > q_\text{el} > q_\text{th}$ & $q_\text{el} > q >q_\text{th}$& $q_\text{el} > q_\text{th}>q$\\
\hline
$C_{\varphi\varphi\varphi\varphi}^R/Y_0$ &   $\frac{1}{1-\nu_0^2}$  & $\frac{1}{1-\nu_0^2}$  & $\sim \left(\frac{q_\varphi}{q_\text{el}}\right)\Omega_{\varphi\varphi\varphi\varphi}\left(q_\varphi/(q_\text{el}q_z)^{1/2}\right) $\\
$C_{\varphi\varphi zz}^R/Y_0$ &   $\frac{\nu_0}{1-\nu_0^2}$  & $\frac{\nu_0}{1-\nu_0^2}$  & $\sim \left(\frac{q_\varphi}{q_\text{el}}\right)\Omega_{\varphi\varphi zz}\left(q_\varphi/(q_\text{el}q_z)^{1/2}\right) $\\
$C_{\varphi z \varphi z}^R/Y_0$ &   $\frac{1}{2(1+\nu_0)}$  & $\frac{1}{2(1+\nu_0)}$  & constant\\
$C_{z z z z}^R/Y_0$ &   $\frac{1}{1-\nu_0^2}$  & $\frac{1}{1-\nu_0^2}$  & constant\\
$Y_{\varphi\varphi}^R/Y_0$ &   $1$  & $1$  & $\sim \left(\frac{q_\varphi}{q_\text{el}}\right)\Omega_{Y_{\varphi\varphi}}\left(q_\varphi/(q_\text{el}q_z)^{1/2}\right)$\\
$Y_{zz}^R/Y_0$ &   $1$  & $1$  & constant\\
$B_{\varphi\varphi\varphi\varphi}^R/\kappa_0$ &   $1$  & $1$ & constant\\
\hline
\end{tabular}
\end{table}

Note that the scaling exponents in the regime $q< \min\{q_\text{th},q_\text{el}\}$ hold to all orders of perturbation theory. This is because of the following reason. Since the parameters in the radial correlation function remain constant due to the irrelevance of all anharmonic terms in the effective free enegy $F_\text{eff}$, the scaling exponents of the in-plane moduli are obtained using simple power counting from self-consistent equations like Eqs.~\ref{eq:C1111Recursion} and~\ref{eq:C2222Recursion}. This remains the same to all orders in perturbation~\cite{RadzihovskyToner}. 

\section{Comparison with molecular dynamics simulations}\label{sec:MD}
To test our results tabulated in Tables~\ref{tab:Scaling1} and~\ref{tab:Scaling2}, we compared with molecular dynamics (MD) simulations. Instead of using a fully atomistic model to simulate a nanotube, we used a convenient coarse-grained discrete model made of a triangular lattice of point masses with nearest neighbors connected by harmonic springs (see Fig.~\ref{fig:figure1}(a)). Then the stretching part in the free energy in Eq.~\ref{eq:FreeEnergy} can be modeled by assigning the equilibrium length of the springs to be $a_0$ and a spring constant $K_b$:
\begin{equation}
    F_\text{stretch} = \frac{1}{2}K_b \sum_{\langle i,j \rangle} (|\mathbf{x}_i - \mathbf{x}_j|-a_0)^2,
\end{equation}
where $\mathbf{x}_i$ is the position of $i^\text{\tiny th}$ mass and the sum is over nearest neighbor point masses $i$ and $j$. The energy cost of bending in Eq.~\ref{eq:FreeEnergy} can be modeled as~\cite{Seung,BowickNumerics}:
\begin{equation}
    F_\text{bending} = K_d \sum_{( \alpha, \beta)} (1+\cos(\theta_{\alpha\beta} -\theta_{\alpha\beta}^0)),
\end{equation}
where $\theta_{\alpha\beta}$ is the angle between two adjacent triangles $\alpha$ and $\beta$ as shown in Fig.~\ref{fig:figure1}(a) and $\theta_{\alpha\beta}^0$ is the value of this angle at minimum bending energy configuration. Note that $\theta_{\alpha\beta}^0$ depends on the curvature of the nanotube and fineness of the discretization. The parameters $K_b$ and $K_d$ are related to the continuum material parameters as follows~\cite{Seung}:
\begin{equation}
    Y_0 = \frac{2}{\sqrt{3}}K_b, \lambda_0 = \mu_0 = \frac{\sqrt{3}}{4} K_b, \nu_0 = \frac{1}{3}, \kappa_0 = \frac{\sqrt{3}}{2}K_d.
\end{equation}
The simulations were done with LAMMPS package~\cite{lammps,lammpsweb}. As will be discussed later, the simulations were done in isobaric-isothermal (NPT) or canonical (NVT) ensemble. Temperature and pressure were controlled using Nos\'e-Hoover type thermostat and barostat~\cite{Tuckerman}. The parameters $K_d$, $K_b$, $T$, the aspect ratio $L/(2\pi R)$ and number of point masses were varied to probe different scaling regimes. The time steps were chosen to be one tenth of the smallest of the characteristic time scales of the system:
\begin{equation}
    \tau_T = a \sqrt{\frac{m}{k_BT}}, \tau_{b} = \sqrt{\frac{m}{K_b}}, \tau_{d} = a_0 \sqrt{\frac{m}{K_d}},
\end{equation}
where $m$ is the mass of each point mass, and $\tau_T$ is characteristic time a point mass takes to cover one atomic length at thermal velocity, $\tau_b$ is characteristic time of the spring-mass system, $\tau_d$ is characteristic time of the dihedral bond-mass system. A simulation generally ran for approximately $1.6 \text{x} 10^8 - 10^9$ time steps. For each simulation, the equilibration was checked using autocorrelation time of different parameters such as radial fluctuations, length of the shell, etc.

First, simulations were done with periodic boundary condition along the axial direction and the simulation box was allowed to change its size in the axial direction maintaining zero pressure condition so that the cylindrical shell could fluctuate freely. From these simulations, the radial displacements were calculated. The Fourier transform of the correlation function $\langle |\tilde{h}(q_\varphi, q_z = 0)|^2\rangle$ and $\langle |\tilde{h}(q_\varphi = 0, q_z)|^2\rangle$ of radial displacement are plotted in Fig.~\ref{fig:figure3}. Fig.~\ref{fig:figure3}(a) shows the collapse around $q_\text{th}$. The dotted lines for $\langle |\tilde{h}(q_\varphi, q_z = 0)|^2\rangle$ and $\langle |\tilde{h}(q_\varphi = 0, q_z)|^2\rangle$ coincide with each other in the region $q>q_\text{el}$ for each parameter set. This is because in this regime the effect of anisotropic curvature is negligible in this regime. Furthermore, in this regime, we see that the correlation function goes as $\sim q^{-4}$ when $q>q_\text{th}$ and $\sim q^{-3.2}$ when $q<q_\text{th}$. This is because in the former case the effect of the anharmonic terms are not important and the system is in the harmonic regime, whereas in the latter case the anharmonic terms are important and since $q>q_\text{el}$ we see the exponent $-4+\eta = -3.2$. However, in the regime $q<q_\text{el}$ they diverge from each other. To understand this regime better, scaling collapse around $q_\text{el}$ is done in Fig.~\ref{fig:figure3}(b) keeping $q_\text{el}\leq q_\text{th}$ for all simulations. Again, for $q>q_\text{el}$, $\langle |\tilde{h}(q_\varphi, q_z = 0)|^2\rangle$ and $\langle |\tilde{h}(q_\varphi = 0, q_z)|^2\rangle$ coincide with each other and scale as $\sim q^{-3.2}$ because here $q_\text{el} < q < q_\text{th}$. However, for $q < q_\text{el} <  q_\text{th}$, we see new scaling laws. In the $\varphi$ direction the correlation function scales as $\sim q_\varphi^{-4}$,  whereas in the $z$ direction the correlation function scales as $\sim q_z^{-1/2}$. These observations can be justified using the Green's function in Eq.~\ref{eq:GreenFunctionRN} and Table~\ref{tab:Scaling1} in the following way:
\begin{subequations}
\begin{equation}
    \langle |\tilde{h}(q_\varphi, q_z = 0)|^2\rangle = \frac{k_BT/A}{B_{\varphi\varphi\varphi\varphi}^R(\mbf{q})q_\varphi^4} = \frac{k_BT/A}{\kappa_R(\mbf{q})q_\varphi^4} \approx \frac{k_BT/A}{\kappa_0 (q_\text{el}/q_\text{th})^{-\eta}q_\varphi^4} \text{ for } q_\varphi<q_\text{el}<q_\text{th},
\end{equation}
\begin{equation}
    \langle |\tilde{h}(q_\varphi = 0, q_z)|^2\rangle = \frac{k_BT/A}{B_{zzzz}^R(\mbf{q})q_z^4+\frac{Y_{\varphi\varphi}^R(\mbf{q})}{R^2}} \approx \frac{k_BT/A}{Y_{\varphi\varphi}^R(\mbf{q})/R^2} \approx \frac{k_BT/A}{Y_0(q_\text{el}/q_\text{th})^{\eta_u}(q_z/q_\text{el})^{1/2}/R^2} \text{ for } q_z<q_\text{el}<q_\text{th}.
\end{equation}
\end{subequations}
In total, both panels of Fig.~\ref{fig:figure3} can be summarized using the following equations:
\begin{figure}[h!]
  \centering
  \includegraphics{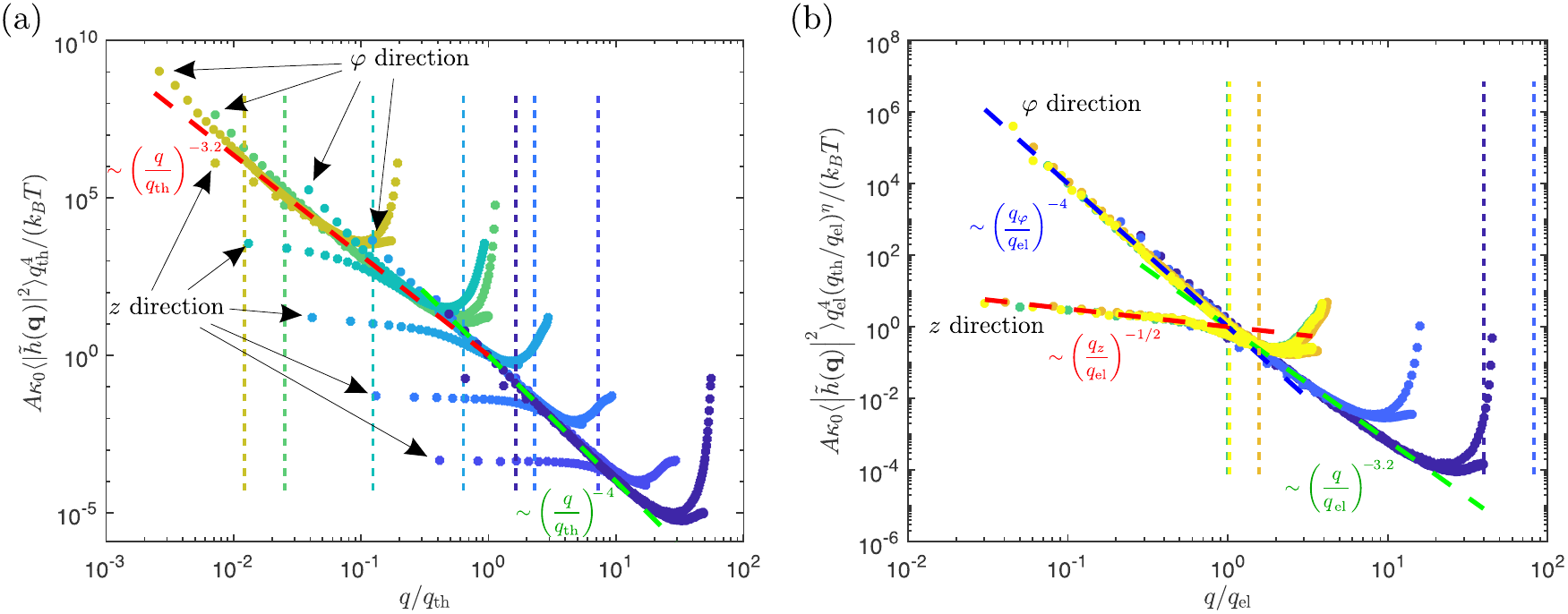}
  \caption{Scaling collapse for radial displacement correlation function $\langle |\tilde{h}(\mbf{q})|^2 \rangle$ for molecular dynamics simulations with zero pressure condition in the axial direction. (a)~Correlation function $\langle |\tilde{h}(q_\varphi, q_z = 0)|^2\rangle$ and $\langle |\tilde{h}(q_\varphi = 0, q_z)|^2\rangle$ collapsed around $q_\text{th}$ plotted in dots. For a single parameter set, the same color was used to plot the correlation function in both $\varphi$ and $z$ direction. The dotted lines of different colors correspond to different parameter sets. The vertical dashed lines of different colors show $q_\text{el}/q_\text{th}$ corresponding to the dotted curves of same color. The red and green slanted dashed lines are $(q_\varphi/q_\text{th})^{-3.2}$ and $(q_\varphi/q_\text{th})^{-4}$ respectively. (b)~Correlation functions $\langle |\tilde{h}(q_\varphi, q_z = 0)|^2\rangle$ and $\langle |\tilde{h}(q_\varphi = 0, q_z)|^2\rangle$ collapsed around $q_\text{el}$ plotted in dots. For a single parameter set, the same color was used to plot the correlation function in both $\varphi$ and $z$ direction. The vertical dashed lines of different colors show $q_\text{th}/q_{el}$ corresponding to the dotted curves of same color. The red, blue and green slanted dashed lines are $(q_z/q_\text{th})^{-1/2}$, $(q_\varphi/q_\text{th})^{-1/2}$ and $(q/q_\text{th})^{-3.2}$ respectively. In both panels, the curling up of the tails of the simulation curves corresponds to wave vectors close to the edge of the first Brillouin zone. }
  \label{fig:figure3}
\end{figure}
\begin{subequations}
\begin{equation}
    \langle |\tilde{h}(q_\varphi, q_z = 0)|^2\rangle = \frac{k_BT/A}{B_{\varphi\varphi\varphi\varphi}^R(\mbf{q})q_\varphi^4} \approx \begin{cases}\frac{k_BT/A}{\kappa_0q_\varphi^4},& q_\text{el}<q_\text{th}<q_\varphi\\\frac{k_BT/A}{\kappa_0 (q_\varphi/q_\text{th})^{-\eta}q_\varphi^4},& q_\text{el}<q_\varphi<q_\text{th}\\\frac{k_BT/A}{\kappa_0 (q_\text{el}/q_\text{th})^{-\eta}q_\varphi^4},& q_\varphi<q_\text{el}<q_\text{th}\end{cases},
\end{equation}
\begin{equation}
    \langle |\tilde{h}(q_\varphi = 0, q_z)|^2\rangle = \frac{k_BT/A}{B_{zzzz}^R(\mbf{q})q_z^4+\frac{Y_{\varphi\varphi}^R(\mbf{q})}{R^2}} \approx \begin{cases}\frac{k_BT/A}{\kappa_R(\mbf{q})q_z^4},& q_\text{el}<q_z\\\frac{k_BT/A}{Y_{\varphi\varphi}^R(\mbf{q})/R^2},& q_z<q_\text{el}\end{cases} \approx \begin{cases}\frac{k_BT/A}{\kappa_0q_\varphi^4},& q_\text{el}<q_\text{th}<q_z\\\frac{k_BT/A}{\kappa_0(q_z/q_\text{el})^{-\eta}q_z^4},& q_\text{el}<q_z<q_\text{th}\\\frac{k_BT/A}{Y_0(q_\text{el}/q_\text{th})^{\eta_u}(q_z/q_\text{el})^{1/2}/R^2},& q_z<q_\text{el}<q_\text{th}\end{cases}.
\end{equation}
\end{subequations}
This confirms the nontrivial scaling of $Y_{\varphi\varphi}$ in the regime $q<q_\text{el}<q_\text{th}$ as we predicted in Section~\ref{sec:AnisotropicScaling}.

\begin{figure}[h!]
  \centering
  \includegraphics{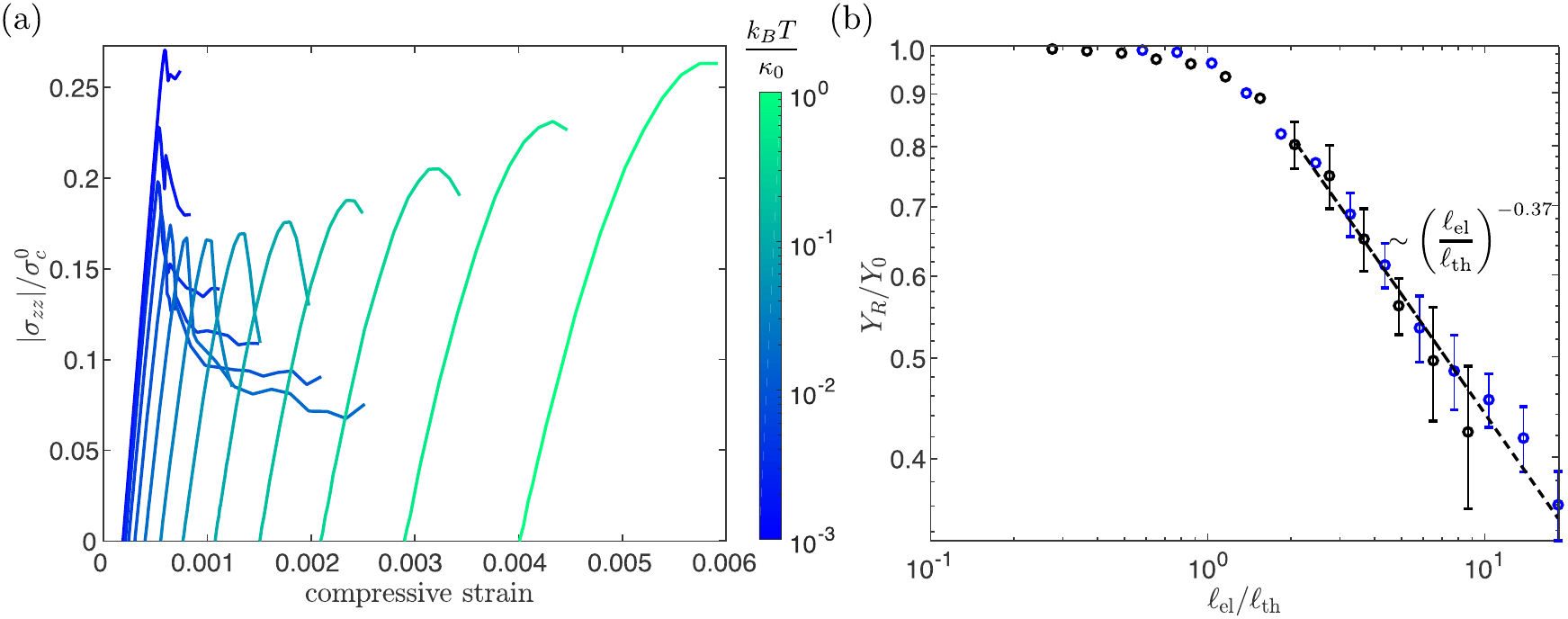}
  \caption{Simulation results from NVT simulations fixing the length of the cylindrical shell. (a)~Stress vs. strain curves for a fixed parameter set $\gamma = 2\times 10^6$, $L/2\pi R = 2$, $2\pi R/a_0 = 48$ at different temperatures (the colorbar shows $k_B/\kappa_0$). The vertical axis of the plot is $|\sigma_{zz}|/\sigma_c^0$, where $\sigma_c^0 = 2\sqrt{Y_0 \kappa_0}/R$ is the critical buckling load for classical cylindrical shell~\cite{Timoshenko}. The maximum of the curve at each temperature is the critical buckling load for molecular dynamics, which is smaller than the classical buckling load $\sigma_c^0$ because of the discrete nature and corresponding non-ideality of our model. (b)~Young's modulus extracted from the slope of the strain vs. strain curves (example shown in (a)) at zero stress plotted as a function of $\ell_\text{el}/\ell_\text{th}$. The black dots correspond to the parameter set $\gamma = Y_0 R^2/\kappa_0 = 1\times 10^5$, $L/2\pi R = 1$, $2\pi R/a_0 = 48$, $L/\ell_\text{el} \approx 35$. The blue dots correspond to the parameter set $\gamma = Y_0 R^2/\kappa_0 = 2\times 10^6$, $L/2\pi R = 2$, $2\pi R/a_0 = 48$, $L/\ell_\text{el} \approx 126$. The black dashed line shows $(\ell_\text{el}/\ell_\text{th})^{-0.37}$.}
  \label{fig:figure4}
\end{figure}

To probe the Young's modulus $Y_{zz}$ in the axial direction, we performed simulations changing the length of the box in steps and at each step letting the shell equilibrate under thermal fluctuation. This ensemble is canonical (NVT) since we fixed the volume of the system at each step. Then, at each value of box length we recorded the pressure of the box in the axial direction. From that, we extracted the normal stress in the axial direction $\sigma_{zz}$ (to get the stress from the pressure, we multiply the pressure with the area of the wall and divide by the perimeter of the nanotube) and plotted the average value of that as a function of strain (defined as the relative change of length from the size of the box at the minimum energy configuration) for different values of temperature $T$ in Fig.~\ref{fig:figure4}(a). Note that in the figure, the strain at which the average stress is zero is different for different temperatures. This is because under thermal fluctuations, the shell shrinks (see Eq.~\ref{eq:AxialShrink}) in equilibrium (at zero external stress condition on the average).
More importantly, we notice that the slope of the stress vs. strain curves change with temperature. The slope of the stress vs. strain curve at zero stress is defined as the Young's modulus $Y_{zz}$ in the axial direction. We plotted the normalized Young's modulus $Y_{zz}/Y_0$, extracted this way, as a function of $\ell_\text{el}/\ell_\text{th}$ in Fig.~\ref{fig:figure4}(b) for two different parameter sets. The Young's moduli for these two parameter sets decrease with increasing system size but only collapse on top of each other when the horizontal axis is $\ell_\text{el}/\ell_\text{th}$ in fig.~\ref{fig:figure4}(b). This implies that $Y_{zz}$ stops renormalizing at the elastic length scale $\ell_\text{el}$ confirming Eq.~\ref{eq:Yzz}. Furthermore, from Fig.~\ref{fig:figure4}(b), we see that the $Y_{zz}^R$ scales as $\sim (\ell_\text{el}/\ell_\text{th})^{-0.37} = (\ell_\text{el}/\ell_\text{th})^{-\eta_u}$ confirming the scaling law in the regime $\ell_\text{th} < \ell < \ell_\text{el}$.
\section{Conclusion}\label{sec:Conclusion}
We have studied the mechanical properties of thermally fluctuating nanotubes. We have shown that the presence of length scales $\ell_\text{th}$ and $\ell_\text{el}$ gives different scaling regimes. In particular, at scales larger than both $\ell_\text{th}$ and $\ell_\text{el}$, we find that the moduli become anisotropic. Moreover, in this regime, we obtain new scaling exponents for the elastic moduli. For system sizes falling between these two lenth scales and with $\ell_\text{th} < \ell_\text{el}$, we recover the same scaling exponents as in the case of isotropic flat solid membranes. We also note that these scaling results have been obtained previously in tubules \cite{RadzihovskyToner} as well as membranes under uni-axial tension \cite{bahri2022mechanical}. It would be of interest to find other physical scenarios that exhibit similar scaling behaviors.

One immediate extension of this work can be to study the effect of thermal fluctuations on the critical axial buckling load. We see from Fig.~\ref{fig:figure4}(a), the critical buckling load initially reduces with increasing temperature (see the bluer curves), but starts increasing (see the greener curves) with increasing temperature. We anticipate that the initial dip is due to the thermal activation over the energy barrier crossing to the buckled state, but the later increase is due to renormalization of elastic moduli. However, deeper investigation is required for better understanding.

The theory presented here is limited to shells with length $L$ to circumference $2\pi R$ with a ratio of order 1. However, shells with $L \gg 2\pi R$ or $L \ll 2\pi R$ may also be importance. In both cases, one can, in principle, integrate out the Fourier modes in the shorter direction and study the effectively 1-dimensional system. In the case $L \gg 2\pi R$, we expect the shell to behave like a polymer chain~\cite{de1979scaling} and perform a random walk when the length is larger than its persistence length. In the case $L \ll 2\pi R$, we expect the shell to behave like an elastic ring which also performs a random walk beyond a persistence length~\cite{RabinPanyukov} and can show interesting instabilities under axisymmetric pressure~\cite{Katifori}.

While the results presented here are for single-walled nanotubes, similar studies can also be done with multi-walled nanotubes. Whereas a realistic model with van der Waals interactions between layers may be too difficult for analytical methods, a phenomenological elasticity-like model as described in~\cite{MauriKatsnelson} adapted for cylindrical shells may be more amenable to analytical studies.

\bibliographystyle{ieeetr}
\bibliography{refer.bib}
\end{document}